\documentclass[preprint]{aastex}
\usepackage{natbib,graphicx}
\newcommand{\flux}{\mbox{ergs cm$^{-2}$ s$^{-1}$}}

\slugcomment{Accepted by the Astronomical Journal}

\received{2004 April 20}
\accepted{2004 July 8}

\shorttitle{Optical Catalog of the CLASXS \mbox{X-ray} Sources}
\shortauthors{Steffen et al.}

\begin{document}


\title{An Optical Catalog of the \emph{Chandra} Large Area Synoptic
  \mbox{X-ray} Survey Sources\altaffilmark{1,2,3,4}}

\author{A. T. Steffen\altaffilmark{5},
  A. J. Barger\altaffilmark{5,6,7},
  P. Capak\altaffilmark{7},
  L. L. Cowie\altaffilmark{7},
  R. F. Mushotzky\altaffilmark{8},
  Y. Yang\altaffilmark{9,8}}

\altaffiltext{1}{Some of the data presented herein were obtained at
  the W. M. Keck Observatory, which is operated as a scientific
  partnership among the California Institute of Technology, the
  University of California, and the National Aeronautics and Space
  Administration.  The observatory was made possible by the generous
  financial support of the W. M. Keck Foundation.}
\altaffiltext{2}{Based in part on data collected at the Subaru
  Telescope, which is operated by the National Astronomical
  Observatory of Japan.}
\altaffiltext{3}{Based in part on data collected at Subaru Telescope
  and obtained from the SMOKA science archive at Astronomical Data
  Analysis Center, which are operated by the National Astronomical
  Observatory of Japan.}
\altaffiltext{4}{Based in part on data collected at the WIYN
  Observatory, which is a joint facility of the University of
  Wisconsin, Indiana University, Yale University, and the National
  Optical Astronomy Observatory.}
\altaffiltext{5}{Department of Astronomy, University of Wisconsin at
  Madison, 475 North Charter Street, Madison, WI 53706.}
\altaffiltext{6}{Department of Physics and Astronomy, University of
  Hawaii, 2505 Correa Road, Honolulu, HI 96822.}
\altaffiltext{7}{Institute for Astronomy, University of Hawaii, 2680
  Woodlawn Drive, Honolulu, HI 96822.}
\altaffiltext{8}{Laboratory for High Energy Astrophysics, Goddard
  Space Flight Center, Code 660, NASA, Greenbelt, MD 20770.}
\altaffiltext{9}{Department of Astronomy, University of Maryland,
  College Park, MD 20742.}

\begin{abstract}
  We present photometric and spectroscopic observations of the
  \mbox{X-ray} sources detected in the wide-area, moderately deep
  \emph{Chandra} Large Area Synoptic \mbox{X-ray} Survey of the
  Lockman Hole-Northwest field.  We have $B$, $V$, $R$, $I$, and $z'$
  photometry for 521 (99\%) of the 525 sources in the \mbox{X-ray}
  catalog and spectroscopic redshifts for 271 (52\%), including 20
  stars.  We do not find evidence for redshift groupings of the
  \mbox{X-ray} sources, like those found in the Chandra Deep
  Field surveys, because of the larger solid angle covered by this survey.
  We separate the \mbox{X-ray} sources by optical spectral type and
  examine the colors, apparent and absolute magnitudes, and redshift
  distributions for the broad-line and non-broad-line active galactic
  nuclei.  Combining our wide-area survey with other {\em Chandra\/}
  and {\em XMM-Newton\/} hard \mbox{X-ray} surveys, we find a definite
  lack of luminous, high accretion rate sources at $z<1$, consistent
  with previous observations that showed that supermassive black hole
  growth is dominated at low redshifts by sources with low accretion
  rates.
\end{abstract}


\keywords{X-ray:galaxies --- general}

\section{Introduction}

The $2-8$ keV \mbox{X-ray} extragalactic background light is dominated
by sources with fluxes around $f_{2-8~{\rm keV}} \simeq 10^{-14}$
\flux\ \citep{cowie02}.  To obtain a more complete picture of the
history of \mbox{X-ray} production in active galactic nuclei (AGNs),
these sources, which have surface densities of only a few hundred
deg$^{-2}$, need to be studied in more detail.  Since the Chandra Deep
Field (CDF) surveys (CDF-N, \citealt{brandt01} and
\citealt{alexander03}; CDF-S, \citealt{giacconi02}) cover too little
solid angle (0.12 and 0.11 deg$^{2}$, respectively) to provide a large
sample of such sources, we have undertaken a wide-area (0.4
deg$^{2}$), contiguous \emph{Chandra} survey of the Lockman Hole --
Northwest field.  The \mbox{X-ray} properties of the sources in our
survey --- called the \emph{Chandra} Large Area Synoptic X-ray Survey,
or CLASXS --- are described in the companion paper by \citet{yang04};
hereafter, Y04.

We designed CLASXS to sample a large, contiguous solid angle, while
remaining sensitive enough to measure sources $2-3$ times fainter than
the observed break in the $2-8$~keV log $N-$log $S$ distribution
\citep[e.g.,][]{mushotzky00,campana01,brandt01,hasinger01,baldi02,rosati02,stern02,alexander03,harrison03,kim04,wang04}.
We chose a survey size of about 0.4 deg$^{2}$ based on \emph{ASCA}
observations, which show a 6\% rms variance for hard \mbox{X-ray}
intensities on a scale of $0.5$ deg$^{2}$ \citep{kushino02}.  This
variance is believed to be responsible for the discrepancies in the
number counts among the ultradeep \emph{Chandra} surveys
\citep{cowie02}.  With CLASXS, we are able to examine the large scale
structure of the \mbox{X-ray} sources \citep{yang03}, something that
cannot be done by combining random, non-contiguous \emph{Chandra}
pointings.  CLASXS is the essential step between the ultradeep, narrow
{\it Chandra} surveys and wide-field, shallow surveys, such as those
obtained by {\em ROSAT} and {\em ASCA}.

In this paper, we present photometric and spectroscopic data of the
optical counterparts to the CLASXS sources.  We briefly describe the
\mbox{X-ray} observations in \S~2. (A full discussion may be found in
Y04.)  We discuss our photometric observations in \S~3,
our spectroscopic data in \S~4, and the photometric and spectroscopic
properties of the CLASXS sources in \S\S~5 and 6, respectively.  In
\S~7, we present the colors and luminosities of the CLASXS sources.  In
\S\S~8 and 9, respectively, we discuss our results and summarize our
main conclusions.  We use J2000 coordinates and the cosmological
parameters $H_{0} = 65$ km s$^{-1}$ Mpc$^{-1}$, $\Omega_{M} =
\case{1}{3}$, and $\Omega_{\Lambda} = \case{2}{3}$.

\section{X-ray Imaging}

To minimize the effects of Galactic attenuation, we chose to position
our CLASXS survey in the Lockman Hole (specifically, the Lockman Hole
- Northwest, one of two {\em Infrared Space Observatory} [{\em ISO}]
fields observed by \citealt{kawara04}), which has the lowest
integrated \ion{H}{1} Galactic column density in the sky \citep[N$_{H}
= 5.7 \times 10^{19}$ cm$^{-2}$;][]{lockman86}.  \citet*{kappes03}
find an additional ionized hydrogen component that comprises $20-50\%$
of the attenuating material; however, limitations on their data
prevent a calculation of the precise ionization fraction.  Regardless,
the total hydrogen content is too low to significantly affect the
\mbox{X-ray} properties.  For reference, we note that our field is
located $\sim2.5\deg$ northwest of the \emph{ROSAT} Lockman Hole field
(\citealt{hasinger98}), which was also observed by \emph{ASCA}
(\citealt{ishisaki01}), \emph{BeppoSAX} (\citealt{giommi00}),
\emph{Chandra} (\citealt{lehmann02}), and \emph{XMM-Newton}
(\citealt{hasinger01}). 

CLASXS consists of nine contiguous \emph{Chandra} ACIS-I exposures
approximately arranged in a $3 \times 3$ grid centered at
$(\alpha,\delta)_{J2000} =$ (10:34:02.1, +57:46:25.0).  The aim points
are separated by about $10\arcmin$.  The central field has a longer
integration (70~ks) than the surrounding eight 40~ks fields.  The
total field coverage is $0.36$ deg$^{2}$, with 70~ks aim point flux
limits (Y04) of $5 \times 10^{-16}$ \flux~ (soft; $0.4-2.0$ keV) and
$3 \times 10^{-15}$ \flux~ (hard; $2-8$~keV).  We have detected 525
independent \mbox{X-ray} sources in at least one of the three
\mbox{X-ray} bands.  In Y04, the \mbox{X-ray} fluxes for the CLASXS
sources were calculated from the observed \mbox{X-ray} counts using
individual power-law indicies that were based on the ratio of the
hard-to-soft \mbox{X-ray} counts (i.e. hardness ratios).  Here we have
converted the $0.4-2$~keV Y04 fluxes into $0.5 - 2.0$~keV fluxes using
a power-law spectral energy distribution with a photon index of
$\Gamma = 1.8$ and a column density of N$_{H}=5.7 \times 10^{19}$
cm$^{-2}$.  If a source is not detected in all of the \mbox{X-ray}
bands, then an upper limit for the flux in any undetected \mbox{X-ray}
band is calculated by measuring the flux at the source position (as
measured in the detected band) using a circular aperture and removing
the local background signal.  If the local background flux is larger
than the flux measured at the source position, then the background
flux is used as an upper limit for the source.  Additional information
regarding the CLASXS \mbox{X-ray} catalog, including reduction and
analysis procedures, can be found in Y04.

All 525 CLASXS \mbox{X-ray} sources from Y04 are listed in
Table~\ref{main_table} in the Appendix, which also gives the 
optical magnitudes and spectroscopic redshifts, when available
(see \S\S~3 and 4).

\section{Optical Imaging}

We obtained deep broadband Johnson \emph{B} and \emph{V}, Cousins
\emph{R} and \emph{I}, and Sloan $z'$ observations reaching $2\sigma$
AB magnitudes of $27.8$, $27.5$, $27.9$, $26.4$, and $26.2$,
respectively, with Suprime-Cam \citep{Miyazaki02} on the Subaru 8.2~m
telescope.  Using the Canada-France-Hawaii Telescope (CFHT) with the
CFH12K camera, we obtained wider field images in broadband Johnson
\emph{B}, Cousins \emph{R}, and CFHT $z'$ reaching $2\sigma$ AB
magnitudes of $27.6$, $27.9$, and $26.3$.  Table~\ref{data_details1}
summarizes the telescopes, dates of the observations, and the
integration times.  Table~\ref{data_details2} summarizes the average
seeing, $2\sigma$ limits, and areas of each image.  The images and the
optical source catalog are presented in \citet{capak04b}, in which a
more extensive discussion of the observations and reductions can be
found.  We briefly summarize the Suprime-Cam and CFH12K observations
and reductions in the Appendix.

\subsection{Optical Source Detection}

We performed source detection on the \emph{R}-band image using the
program SExtractor \citep{bertin96}.  To improve faint source
detection and minimize sky noise, the image was smoothed using a
$2\arcsec$ Gaussian kernel.  The detection threshold was set to be
eight contiguous pixels with counts at least $2~\sigma$ above the sky
background.  We measured the optical magnitude of each source with the
IDL routine APER, using a 3\arcsec\ diameter aperture centered on the
optical counterpart, if detected, or on the \mbox{X-ray} position if not.
The magnitudes of bright sources that are
significantly larger than the 3\arcsec ~aperture are underestimated,
as are $R<20$ sources, which are typically saturated in our images.

Table~\ref{main_table} in the Appendix gives the optical properties of
the CLASXS X-ray sources.  The CLASXS source number (from Y04) is given in
column (1).  The positions for the detected optical counterparts are
given in columns (5)-(6).  The aperture-corrected $B$, $V$, $R$, $I$,
and $z'$ magnitudes are given in columns (9)-(13).  Column (14) gives
the measured redshifts of identified sources.  In this column, we
labeled spectroscopically identified stars ``star''.  If we
spectroscopically observed a source but did not identify it, then we
labeled the source ``obs''.

The aperture correction was calculated by examining the
curve of growth for isolated, moderately bright ($R = 20-26$) sources.
The 3\arcsec\ aperture corrections for the Subaru (CFHT) images are
$-0.17$ ($-0.30$), $-0.22$, $-0.11$ ($-0.15$), $-0.33$, and $-0.13$
($-0.22$), for the \emph{B}, \emph{V}, \emph{R}, \emph{I}, and $z'$
bands, respectively.  Sources that fall outside of the field of view
of a given filter are left blank.  If a source falls within the
field of view but is not detected, then it is assigned the $2~\sigma$
magnitude limit for that filter and is considered an upper limit.

\subsection{Optical Counterpart Identifications of the CLASXS 
\mbox{X-ray} Sources}

{\it Chandra}'s cylindrical mirrors were designed to be the smoothest
mirrors ever constructed in an effort to obtain the highest resolution
X-ray images ever ($\sim0.5\arcsec$ at the aim point), and thus the
most accurate positions for the sources. This unprecedented
\mbox{X-ray} positional accuracy allows the secure identification of
counterparts at other wavelengths.  The cylindrical mirror design,
which is required for the grazing incidence optics needed to focus
\mbox{X-rays}, introduces an aberration that increases the point
spread function (PSF) of an image with an off-axis angle.  For example,
{\em Chandra}'s PSF at an off-axis angle of $6\arcmin$ ($12\arcmin$)
from the aim point is $\sim 2.5\arcsec$ ($\sim 9\arcsec$).  This
increasing \mbox{X-ray} PSF size with off-axis angle makes it
difficult to identify optical counterparts using a fixed search
radius.  If too small a radius is used, then counterparts for
\mbox{X-ray} sources with large PSFs may be missed.  If too large a
radius is used, then it will negate \emph{Chandra}'s excellent
resolution and ability to securely identify \mbox{X-ray} sources with
no detectable optical counterpart.  Because of this, we divide the
CLASXS sample into two groups using the radial distance from the aim
point ($R_{aim}$; col [4] in Table~\ref{main_table}).  A 2\arcsec\ 
search radius is used for sources with $R_{aim} < 6\arcmin$, and
3\arcsec\ for sources with $R_{aim} \geq 6\arcmin$.  We use the
smallest aim point distance if a source is detected in more than one
of the CLASXS pointings.  Since the nine individual CLASXS fields that
make up the larger mosaic seperated by only $10\arcmin$, the majority
of our sources are less than $8\arcmin$ from an aim point of one of
these fields.  Thus, all of our positions are relatively uniformly
determined.  If more than one optical source falls within the
search radius, then we designate the closest optical neighbor as
the counterpart.  Using these criteria, 484 ($92\%$) of the 525
\mbox{X-ray} detected sources in the Y04 catalog have optical
counterparts, and 264 ($50\%$) of these have magnitudes that are
bright enough for straightforward spectroscopic follow-up ($R<24$).

%
%
\begin{figure}[th]
  \epsscale{1.0}
  \plotone{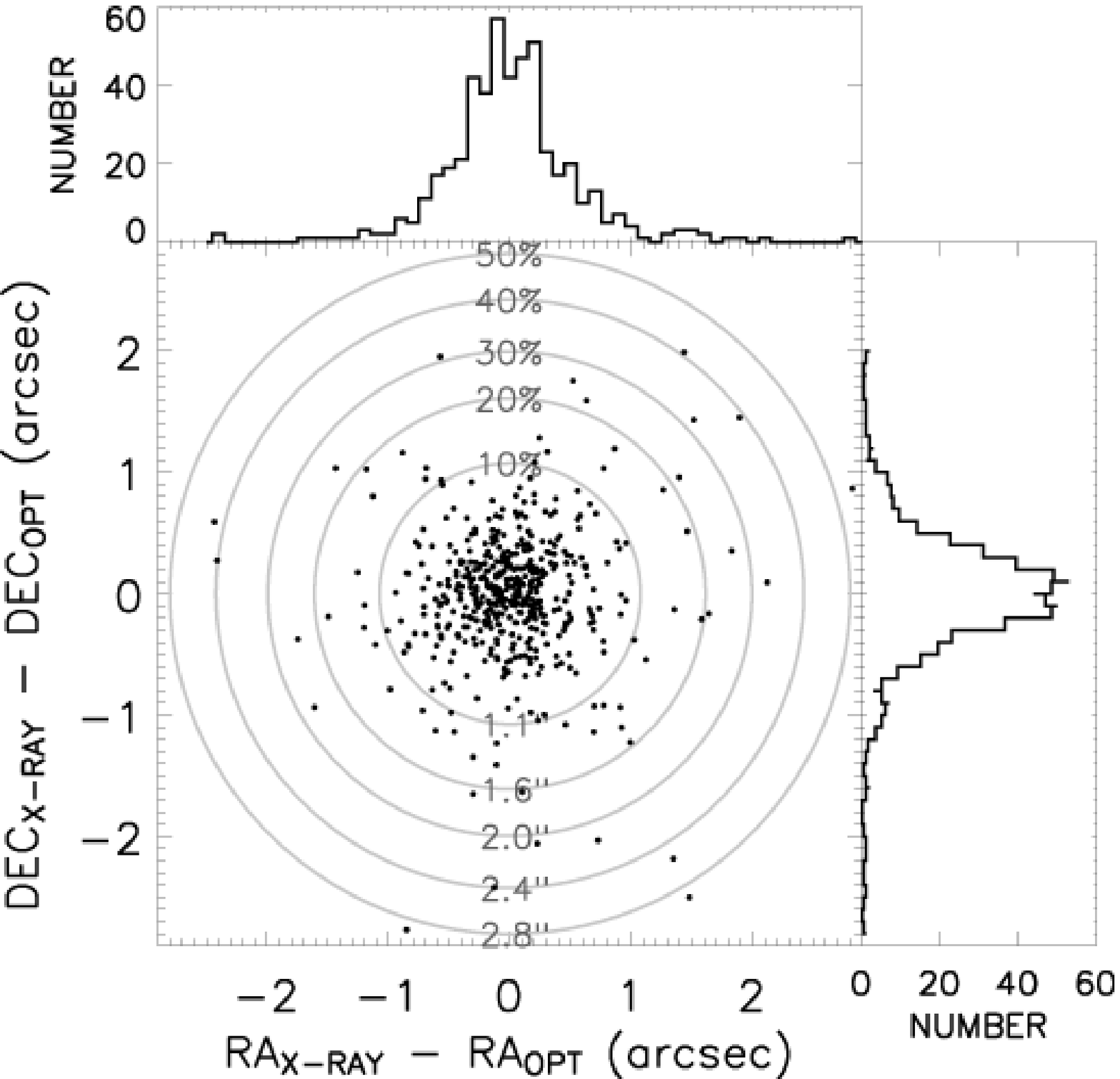}
  \caption{\label{astrometry}(\mbox{X-ray} $-$ optical) astrometric offsets 
    for the 484 CLASXS sources with detected optical counterparts.
    Histograms for the R.A. (decl.) separations are shown on top ({\it
      right}).  The mean values for the R.A. and decl. offsets are
    both $0.0 \pm 0.5$ arcseconds.  Concentric grey circles represent
    the probability of a source with a random R.A. and decl. being
    assigned an optical counterpart.  The probabilities and search
    radii (in arcseconds) are given, respectively, at the top and
    bottom of each circle.}
\end{figure}

In Figure~\ref{astrometry}, we show the (\mbox{X-ray} $-$ optical)
astrometric offsets for the CLASXS sources.  Histograms for the right
ascension and declination offsets are shown above and to the right,
respectively.  The average astrometric discrepancies in both axes are
$0.0 \pm 0.5$ arcseconds.  We ran simulations to examine the
probability of an \mbox{X-ray} source being assigned an incorrect
optical counterpart.  The probability of a chance projection is a
strong function of the limiting optical magnitude in the catalog,
since there are many more sources at the faint end.  The concentric
circles represent the probability of a source with a random right
ascension and declination being assigned an optical counterpart from
the full optical catalog ($R<27.9$).  The inner circle is 10\%, and
the other circles increase outward with 10\% increments.  The majority
of CLASXS sources have (\mbox{X-ray} $-$ optical) separations of less
than $0.5\arcsec$ ($1~\sigma$) for which there is only a 2\%
probability of an incorrect \mbox{X-ray}/optical match.  With our
search radius of $2\arcsec$, we calculate that 30\% of the CLASXS
sources that in fact have no optical counterpart will be assigned an incorrect
optical counterpart.  Almost all of these will be with optically
faint sources.  If we limit the optical sources to those
spectroscopically accessible ($R<24$), we find only an 8\% chance of
an incorrect \mbox{X-ray}/optical overlap using a $2\arcsec$ matching
radius.

We show $R$-band thumbnails for 521 of the 525 CLASXS \mbox{X-ray}
sources in the Y04 catalog in Figure~\ref{thumbnails} in the Appendix,
ordered by increasing right ascension.  Four \mbox{X-ray} sources (192, 318, 328,
and 359) are outside the fields of view of our optical images and are
not included.  Stars are labeled ``star''.  Optically undetected
sources are labeled ``B'' for ``blank''.  Spectroscopically observed
sources that could not be identified are labeled ``obs''.

\section{Spectroscopic Observations}

Optical spectra were obtained using the multi-fiber spectrograph HYDRA
\citep{barden94} on the WIYN 3.5~m telescope for bright ($I<19$)
sources, and with the Deep Extragalactic Imaging Multi-Object
Spectrograph \citep[DEIMOS;][]{faber03} on the 10~m Keck~II telescope
for fainter sources.  Table~\ref{spec_obs_summary} summarizes the telescopes,
instruments, and dates of the spectroscopic observations.

%
%
\begin{figure}[th]
  \epsscale{1.0}
   \includegraphics[angle=90,scale=0.6]{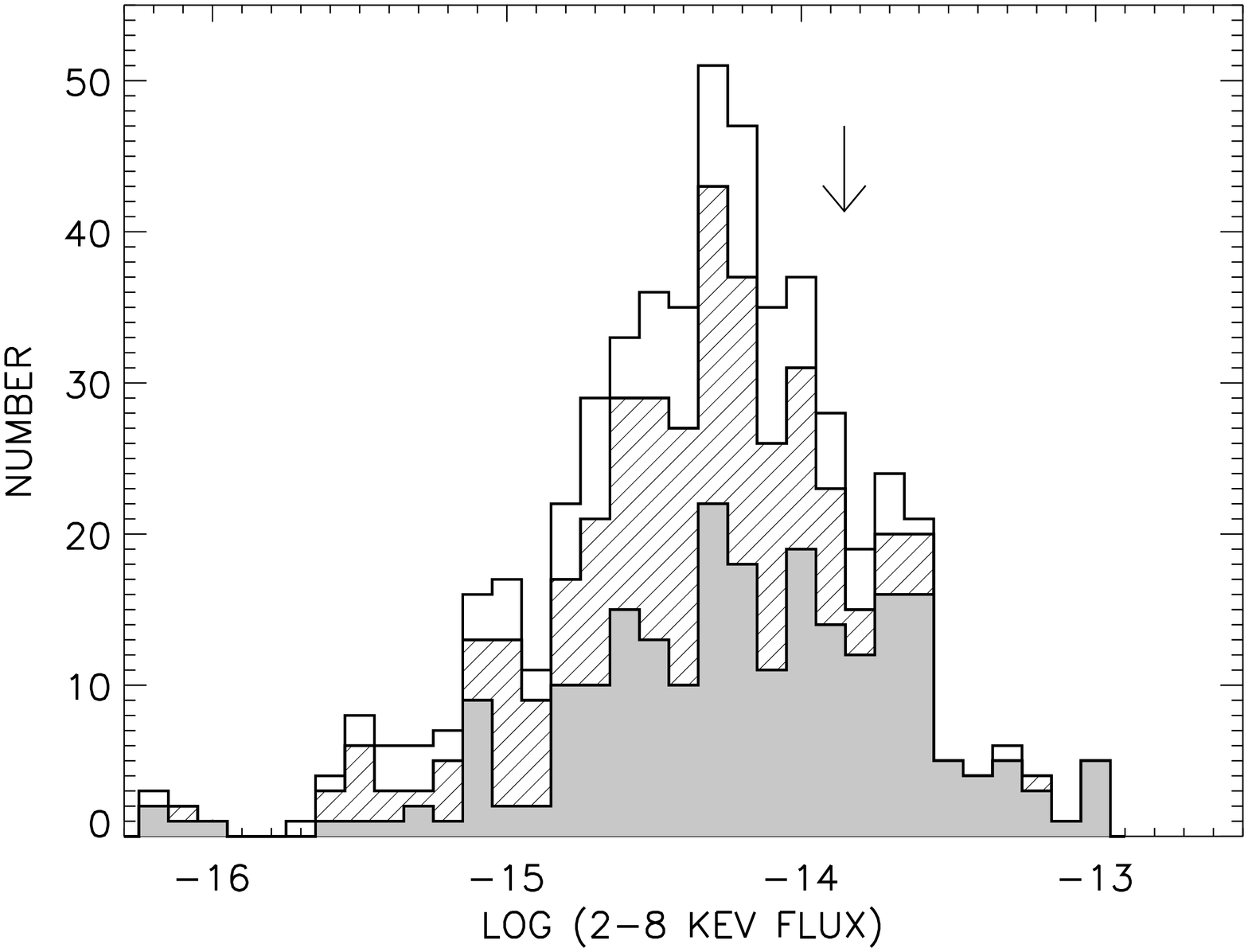}
  \caption{\label{log_flux_hist}$2 - 8$~keV flux distribution for the
    525 CLASXS sources ({\it shaded}, spectroscopic redshifts; {\it
      hatched}, spectroscopically observed sources that could not
    be identified; {\it open}, spectroscopically unobserved
    sources).  The arrow marks the flux where the source contribution to
    the $2-8$ keV~XRB peaks.}
\end{figure}

%
%
\begin{figure*}[t]
  \epsscale{0.6}
  \plotone{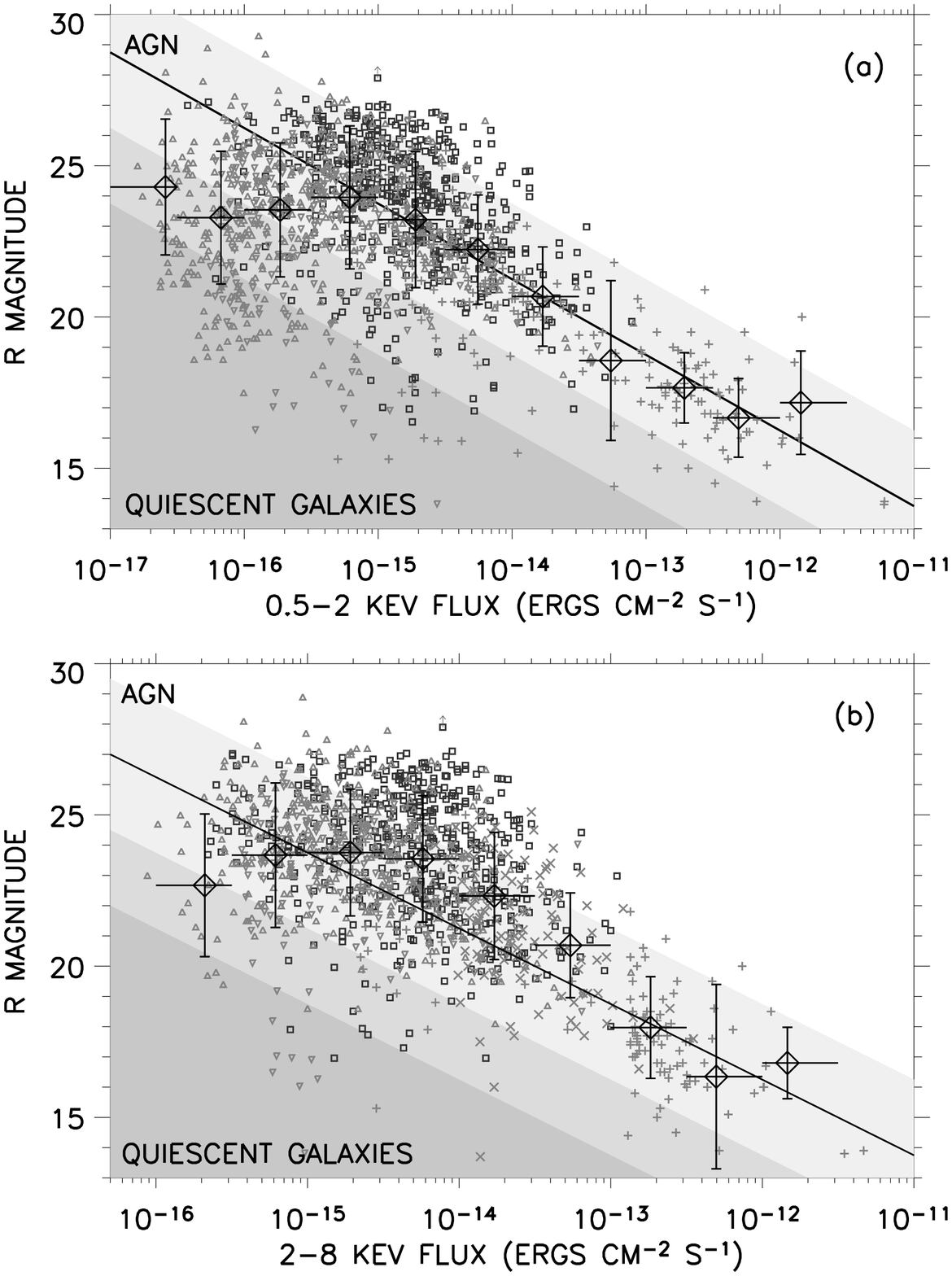}
  \caption{\label{r_mag_vs_sb} ({\it a}) The $R$ magnitude vs. $0.5-2.0$~keV flux 
    for the soft \mbox{X-ray} sources in CLASXS ({\it squares}; Y04),
    the CDF-N ({\it upward-pointing triangles}; Barger et al.\ 2003), the CDF-S
    ({\it downward-pointing triangles}; Szokoly et al.\ 2004), and in combined
    \emph{ROSAT} \citep{lehmann01}, \emph{XMM-Newton}
    \citep{mainieri02}, and \emph{ASCA} \citep{akiyama03} surveys
    ({\it plus signs}). ({\it b}) The $R$ magnitude vs. $2-8$~keV flux for the
    hard \mbox{X-ray} sources (symbols as in ({\it a}); Lehmann et al.
    \emph{ROSAT} sources have no hard \mbox{X-ray} measurements).  In
    ({\it b}) we added the hard \mbox{X-ray} data from the HELLAS2XMM survey
    ({\it crosses}; Fiore et al.\ 2003).  In both ({\it a}) and ({\it b}), the
    solid line shows where the fluxes are equal (i.e.,
    log$(f_{X}/f_{R}) = 0$). The lightly shaded region covers
    log$(f_{X}/f_{R}) = \pm 1$, where AGNs typically reside.
    Quiescent galaxies are usually found in the darkest shaded region
    (log$(f_{X}/f_{R}) < -2$). Median magnitudes and fluxes for the
    \mbox{X-ray} sources from Tables \ref{optical_soft} and
    \ref{optical_hard} are shown as large diamonds with $1\sigma$
    ranges shown as error bars. [{\it A color version of this figure is
    available in the electronic edition of the Astronomical Journal.}]}
\end{figure*}

%
%

For the HYDRA observations, we used a low-resolution grating, 316@7.2,
centered at 7600~\AA, yielding a wavelength coverage of
$4900-10300$~\AA\ with a resolution of 2.64~\AA pixel$^{-1}$.  The red bench
camera was used with the 2\arcsec\ ``red'' HYDRA fibers to maximize
the sensitivity at longer wavelengths.  To obtain a wavelength
solution for each fiber, CuAr comparison lamps were observed in each
HYDRA configuration.  Our HYDRA masks were designed to maximize the
number of optically bright sources in each configuration, while
minimizing the amount of overlap between configurations.  Fibers that
were unable to be placed on a source were assigned to a random sky
location.  We observed 2.7 hrs on two HYDRA configurations in 2001
February, 7.4 hrs on two configurations in 2002 February , and 6.0 hrs on two
configurations in 2002 March.  To remove fiber-to-fiber variations,
on-source observations were alternated with $\pm7.5\arcsec$ ``sky''
exposures taken with the same exposure times.  This effectively
reduced our on-source integration times to 1.3, 3.6, and 3.0 hrs in
2001 February, 2002 February, and 2002 March, respectively.

Reductions were performed using the standard IRAF\footnote{IRAF is
  distributed by the National Optical Astronomy Observatory, which is
  operated by the Association for Research in Astronomy, Inc., under
  cooperative agreement with the National Science Foundation.}
package DOHYDRA.  To optimize sky subtraction, we performed a two-step
process.  In the first step, the DOHYDRA routine was used to create an
average sky spectrum using the fibers assigned to random sky
locations.  This average sky spectrum was then removed from all of the
remaining fibers.  In the offset images, this step effectively removed
all of the sky signal, leaving behind only residuals caused by
differences among the fibers.  To remove these variations, we then
subtracted the residuals present in the sky-subtracted offsets from
the sky-subtracted, on-source spectra.  We found this method to be
very effective at removing the residuals created by fiber-to-fiber
variations with HYDRA.

For the DEIMOS observations, we used the 600 lines mm$^{-1}$ grating,
which yielded a resolution of 3.5~\AA\ and a wavelength coverage of
5300~\AA.  The exact central wavelength depends upon the slit
position, but the average was 7200~\AA.  Each $\sim 1$ hr exposure was
broken into 3 subsets.  In each subset the object was stepped
1.5\arcsec\ in each direction.  The DEIMOS spectroscopic reductions
follow the same procedures used by \citet{cowie96} for LRIS
reductions. The sky contribution was removed by subtracting the median
of the dithered images.  Cosmic rays were removed by registering the
images and using a cosmic ray rejection filter on the combined images.
Geometric distortions were also removed and a profile-weighted
extraction was applied to obtain the spectrum.  Wavelength calibration
was done using a polynomial fit to known sky lines rather than using
calibration lamps.  The spectra were individually inspected and a
redshift was measured only for sources where a robust identification
was possible.  The high-resolution DEIMOS spectra can resolve the
doublet structure of the [\ion{O}{2}]~$\sim 3727$~\AA~line, allowing
spectra to be identified by this doublet alone.

Figure~\ref{log_flux_hist} shows the $2-8$~keV flux distribution for
the CLASXS sources.  The sources that were spectroscopically observed
{\it (hatched)} and spectroscopically identified {\it (solid)} are
indicated.  The arrow marks the flux where the source contribution to
the $2-8$~keV XRB peaks.  Of the 525 CLASXS
sources detected in Y04, we have spectroscopically observed 467
(90\%) and identified 271 (52\%), including 20 stars.

\section{Optical Properties of the \mbox{X-ray} Sources}

Tables~\ref{optical_soft} and \ref{optical_hard} give the median
optical magnitudes for the CLASXS soft and hard \mbox{X-ray} sources,
respectively.  The tables also give, in parentheses, the median
optical magnitudes for the CLASXS sources combined with other
observations from {\it Chandra} \citep{barger03,szokoly04}, {\it
  ROSAT} \citep{lehmann01}, {\it XMM-Newton}
\citep{mainieri02,fiore03}, and {\it ASCA} \citep{akiyama03}.
Figure~\ref{r_mag_vs_sb} shows the \emph{R} magnitudes of the combined
samples versus (a) soft and (b) hard \mbox{X-ray} flux.
Any of the published \mbox{X-ray} catalogs that used
energy ranges for the soft and hard bands other than $0.5-2$~keV and
$2-8$~keV, respectively, were converted to this range using the power
law assumed by the respective authors, or, if none were given, then by
assuming a power law of $\Gamma = 1.8$ and no absorption.  The CLASXS
sources with optical upper limits are denoted by arrows.  CLASXS
sources brighter than $R=20$ are typically saturated in our images, so
the measured magnitudes of these sources may be underestimated.

The shaded regions in Figures~\ref{r_mag_vs_sb}{\it a} and
\ref{r_mag_vs_sb}{\it b} represent typical ranges of log$(f_{X}/f_{R})$ for
different source classes \citep[e.g.,][]{alexander02,bauer02,barger03}.  
The flux in the \emph{R}-band, $f_{R}$, is related to the \emph{R} 
magnitude by log$(f_{R})=-5.5 - 0.4R$ \citep{horn01}.  AGNs typically 
lie in or above the region defined by the loci 
log$(f_{X}/f_{R}) = \pm 1$ {\it (lightest shading)}, while normal 
galaxies lie below log$(f_{X}/f_{R}) < -2$ {\it (darkest shading)}.
Note that the flux limit of the CLASXS survey
($f_{2-8 {\rm keV}} > 3 \times 10^{-15}$ \flux) makes it an AGN
dominated survey.

The median $R$-band magnitudes in each flux bin (see values in
parentheses in Tables~4 and 5) are shown in Figure~\ref{r_mag_vs_sb}
as the large diamonds with $1\sigma$ standard deviations for each flux
bin, denoted by error bars.  The median \mbox{X-ray}/optical flux
ratios for the soft \mbox{X-ray} sources follow the log$(f_{X}/f_{R})
= 0$ loci for fluxes larger than $f_{0.5-2~{\rm keV}} \ge 5 \times
10^{-16}$ \flux, but drop to smaller \mbox{X-ray}/optical ratios at
fainter \mbox{X-ray} fluxes.  As noted by \citealt{barger03}, the
explanation for this is straightforward. At bright soft \mbox{X-ray}
fluxes, the AGNs usually dominate the \mbox {X-ray} and optical light,
so the galaxies follow a constant \mbox{X-ray}/optical flux ratio.  At
fainter \mbox{X-ray} fluxes, the starlight begins to dominate over the
optical AGN contributions, possibly as a result of obscuration, an
increase in the number of intrinsically lower luminosity AGNs, or, at
the faintest fluxes, an increase in the number of galaxies without
AGNs \citep[e.g.][]{horn03}. Since stars contribute less than the AGNs
to the \mbox{X-ray} emission, the result is a drop in the
\mbox{X-ray}/optical flux ratios.  In addition, any intrinsic
obscuration will also attenuate the soft \mbox{X-ray} emission from
the central engine.  For near Compton-thick AGNs, all of the soft
\mbox{X-ray} and optical AGN light will be attenuated.

The $R$-magnitudes shown in Figure~\ref{r_mag_vs_sb} are not nuclear
magnitudes, so they differ from the optical/\mbox{X-ray} studies done of
nearby Seyfert galaxies.  We use a fixed aperture for our photometric
measurements that covers a varying fraction of the optical
counterpart.  This fraction is dependent on the apparent angular size
of the optical source and can vary between the nuclear magnitude for
nearby sources and the total magnitude for high-redshift sources.
This introduces a scatter in the y-axis of Figure~\ref{r_mag_vs_sb},
because a high-redshift source will have a brighter optical magnitude
than its low-redshift twin.  Since most of our sources have small
angular sizes (see Figure~\ref{thumbnails}), our $R$-band magnitudes
can be considered upper limits to the nuclear $R$-band magnitudes.  In
other words, if we could measure only the nuclear magnitudes for the
CLASXS sources, then the observed $R$-band magnitudes would decrease.
This would shift many CLASXS sources away from the Quiescent Galaxies
regime and into the AGN regime.

\subsection{AGN Obscuration}

%
%
\begin{figure*}[hb]
  \epsscale{1.0}
\plotone{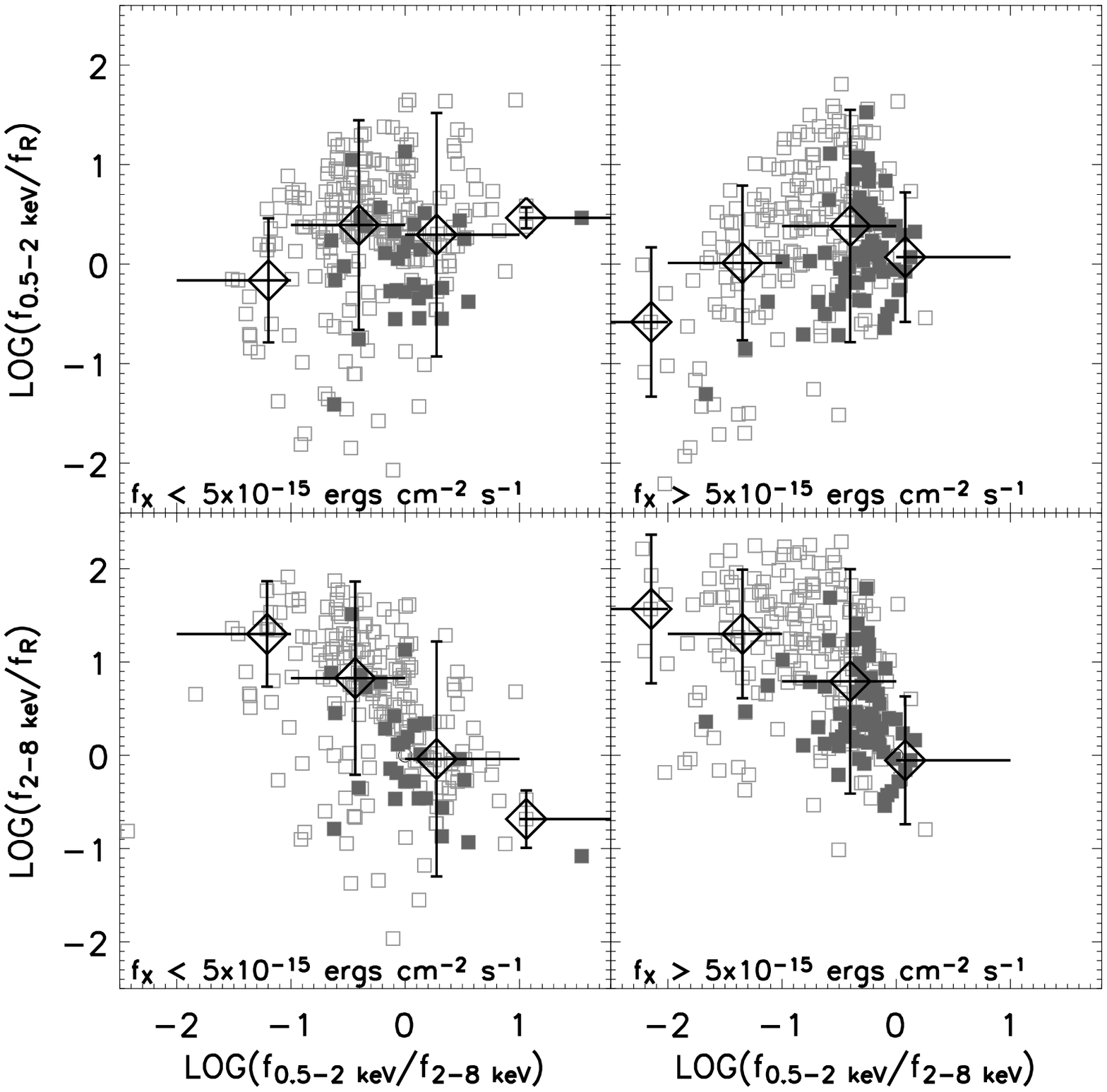}
\caption{\label{fr_vs_hr}Soft and hard \mbox{X-ray}--to--optical flux 
  ratios for the CLASXS soft and hard \mbox{X-ray}--selected sources 
  vs. soft to hard \mbox{X-ray} flux ratio.  The
  samples were further broken into
  two populations: \mbox{X-ray} bright ($f_{2-8~{\rm keV}} > 10^{-15}$
  \flux) and \mbox{X-ray} faint ($f_{2-8~{\rm keV}} < 10^{-15}$
  \flux). Sources identified as stars are not shown.  BLAGNs are shown
  as solid squares and non-BLAGNs as open squares.  The average flux
  ratio is shown as large open diamonds for each logarithmic
  soft-to-hard flux ratio bin.}
\end{figure*}

The presence of an obscuring torus can affect the \mbox{X-ray}/optical
flux ratio of a source, depending on which \mbox{X-ray} band we
examine.  Since a photon's cross-section decreases rapidly with
increasing photon energy, light from the near-infrared to soft \mbox{X-rays}
is typically attenuated in moderately obscured AGNs, while hard \mbox{X-ray}
light penetrates all but the most obscured sources.

Figure~\ref{fr_vs_hr} shows the $0.5-2$~keV to $R$-band and $2-8$~keV
to $R$-band flux ratios versus soft-to-hard \mbox{X-ray} flux ratio
for the CLASXS soft and hard \mbox{X-ray} selected samples.  The
samples were further separated by hard \mbox{X-ray} flux into bright
($f_{2-8~{\rm keV}} > 10^{-15}$ \flux) and faint ($f_{2-8~{\rm keV}} <
10^{-15}$ \flux) categories.  The filled squares denote broad-line
AGNs (BLAGNs).  The large diamonds are the average $\log(f_{X}/f_{R})$
within the soft-to-hard flux ratio ranges of $\log (f_{0.5-2~{\rm
    keV}}/f_{2-8~{\rm keV}}) = (-2,-1)$, $(-1,0)$, $(0,1)$, $(1,2)$,
and $(2,3)$.

We expect that for the $0.5-2$~keV \mbox{X-ray}/optical flux ratio,
the presence of any obscuring material will attenuate both the soft
\mbox{X-ray} and optical photons, making the flux ratio only weakly
dependent on the column density for unobscured to moderately obscured
sources.  Moreover, for highly obscured sources, the AGN contribution
to both the observed soft \mbox{X-ray} and optical light should become
negligible, causing the galaxy to descend into the ``Quiescent
Galaxy'' category.

If we treat the soft-to-hard flux ratio as an approximation to the
line-of-sight column density (column density would decrease with
increasing soft-to-hard \mbox{X-ray} flux ratio), we can see from
Figure~\ref{fr_vs_hr} that the soft \mbox{X-ray}/optical flux ratio
{\it (top row)} does indeed remain relatively constant for
$\log(f_{0.5-2~{\rm keV}}/f_{2-8~{\rm keV}}) > -1$.  Moreover, at
larger column densities, where both the optical and the soft
\mbox{X-ray} light are obscured, there may be a slight drop in the
soft \mbox{X-ray}/optical flux ratio as the AGN contributions to both
become negligible and the optical starlight from the galaxy begins to
dominate.

In contrast, we expect that moderate column densities of material will
not significantly attenuate the hard \mbox{X-ray} photons from an AGN,
so moderately obscured sources will have high hard \mbox{X-ray}/optical flux
ratios.  From Figure~\ref{r_mag_vs_sb} {\it (bottom row)}, we see that
as the column density increases, the AGN's optical light is blocked,
while the hard \mbox{X-rays} can still penetrate the material.  Thus,
the hard \mbox{X-ray}/optical flux ratio increases until the material
becomes opaque to hard \mbox{X-rays}.  At these high column densities, even
the \mbox{X-ray} light is absorbed, and we would expect to see the
\mbox{X-ray}/optical flux ratio begin to drop.  However, near-Compton-thick
sources are very \mbox{X-ray} weak and difficult to detect, and there
may be very few of them in the sample.

The aforementioned trends in the soft and hard \mbox{X-ray}-to-optical
ratios appear to hold for both the \mbox{X-ray} bright and faint CLASXS
sources.  However, there are noticeable differences in the two
populations that can be seen in Figure~\ref{fr_vs_hr}.  There is a
significantly larger population of BLAGNs in the \mbox{X-ray} bright
population than in the \mbox{X-ray} faint population.  This is consistent
with the discovery by \citet{steffen03} that the BLAGNs dominate the
number densities at the higher \mbox{X-ray} luminosities, while the
non-BLAGNs dominate at the lower \mbox{X-ray} luminosities.

\section{Spectroscopic Properties}

\begin{figure}[ht]
  \epsscale{1.0}
   \includegraphics[angle=90,scale=0.6]{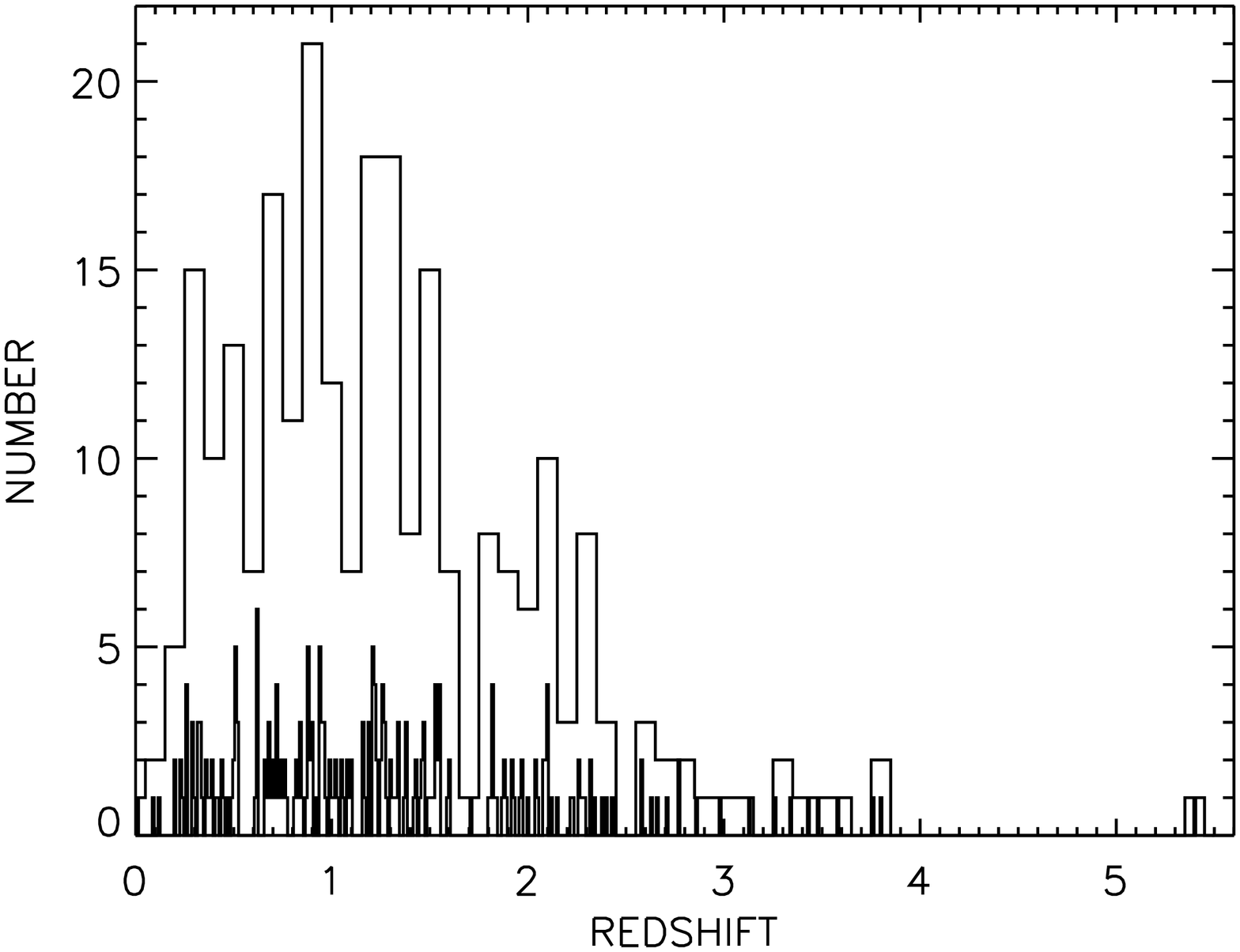}
  \caption{\label{redshift_hist}Redshift distribution for the
    spectroscopically identified CLASXS sources.  All stars and sources
    without measured redshifts have been excluded.  Two different bin
    sizes ($\Delta z = 0.1$ and $\Delta z = 0.01$) are shown.}
\end{figure}

The redshift distribution of the CLASXS sources is shown in
Figure~\ref{redshift_hist}.  The two histograms have $\Delta z = 0.1$
and $\Delta z = 0.01$ width bins.  There are no obvious structures at
$z<1$ in the CLASXS redshift distribution, unlike the apparent
excesses found at $z=0.48$ and $z=0.94$ in the CDF-N sample by
\citet{barger03}, and at $z=0.674$ and $z=0.734$ in the CDF-S sample
by \citet{gilli03} and \citet{szokoly04}.  (There might be a grouping
of CLASXS sources at $z=1.8$, though this is not easily seen in the
figure.)  The differences between the redshift distributions of CLASXS
and the CDFs show that we have indeed sampled a large enough area in
CLASXS to average out the large scale structure.

%
%

Following \citet{cowie04a}, we optically classified sources into
four categories, roughly using the procedure of \citet{sadler02} to
analyze low-redshift radio samples.  We classified sources without any
strong emission lines (EW([\ion{O}{2}])$ < 3$\AA\ or EW(H$\alpha +
$\ion{N}{2} $< 10$\AA) as absorbers; sources with strong Balmer lines
and no broad or high-ionization lines as star formers; sources with
[\ion{Ne}{5}] or \ion{C}{4} lines or strong [\ion{O}{3}]
(EW([\ion{O}{3}] 5007 \AA) $ > 3$ EW(H$\beta$)) as Seyfert galaxies; and,
finally, sources with lines with FWHM $> 1000$ km s$^{-1}$ as BLAGNs.
Table~\ref{source_type} breaks down the identified sources by spectral
type.

%
%
\begin{figure}[bh]
  \epsscale{1.0}
   \includegraphics[angle=90,scale=0.6]{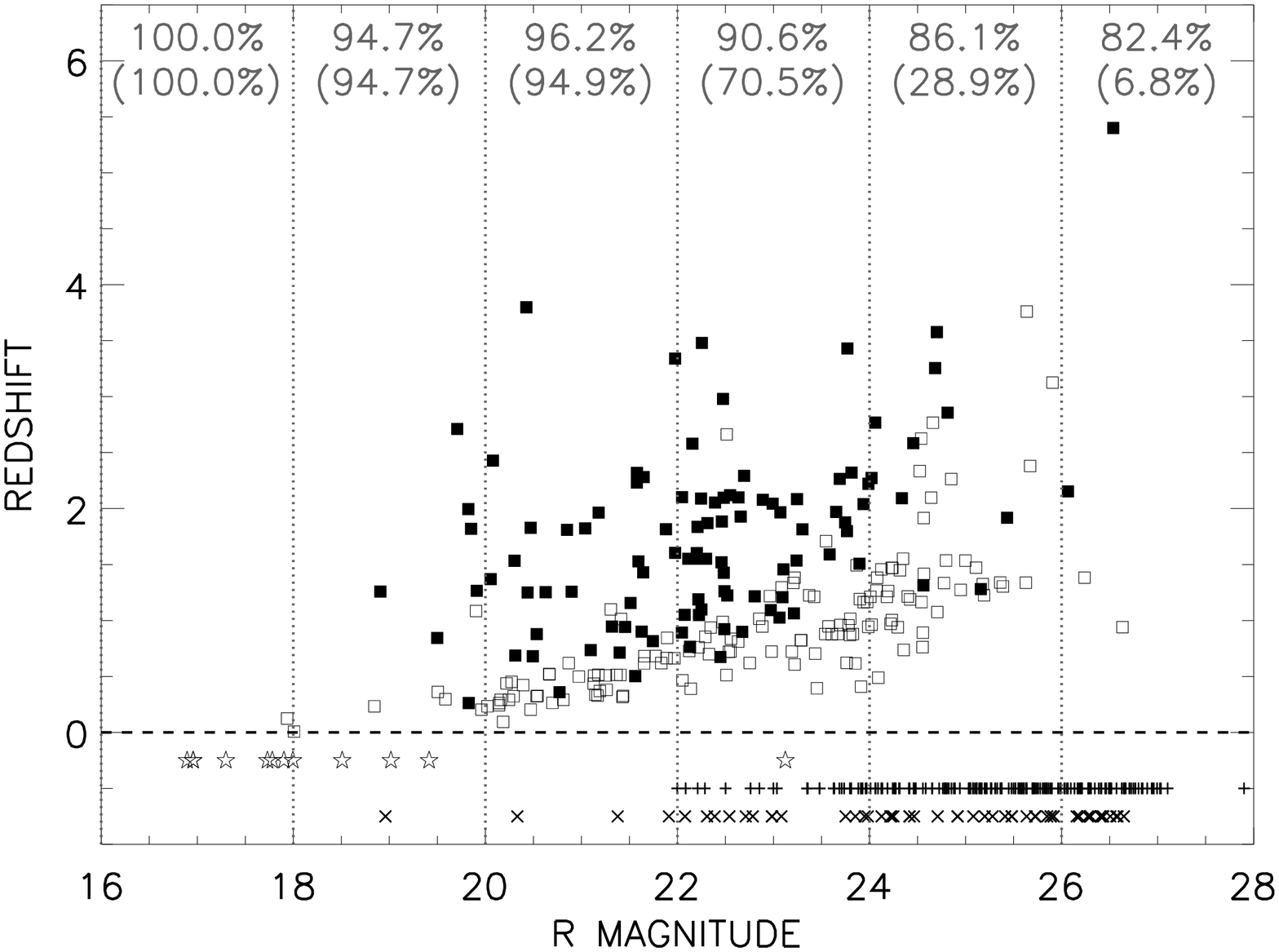}
  \caption{\label{r_mag_vs_redshift} Redshift vs. \emph{R}-band
    magnitude for the CLASXS hard \mbox{X-ray} sample ({\it stars},
    stars placed at $z=-0.25$; {\it plus signs},spectroscopically
    observed but unidentified sources placed at $z=-0.5$; {\it
      crosses},spectroscopically unobserved sources placed at
    $z=-0.75$; {\it filled squares},BLAGNs; {\it open
      squares}---non-BLAGNs).  The graction of the CLASXS sources that
  have been spectroscopically observed (identified) for each bin of
  $\Delta R = 2$ mag ({\it dotted lines}) is shown across the top of
  the figure.}
\end{figure}

\begin{figure}[hb]
  \epsscale{1.0}
   \includegraphics[angle=90,scale=0.6]{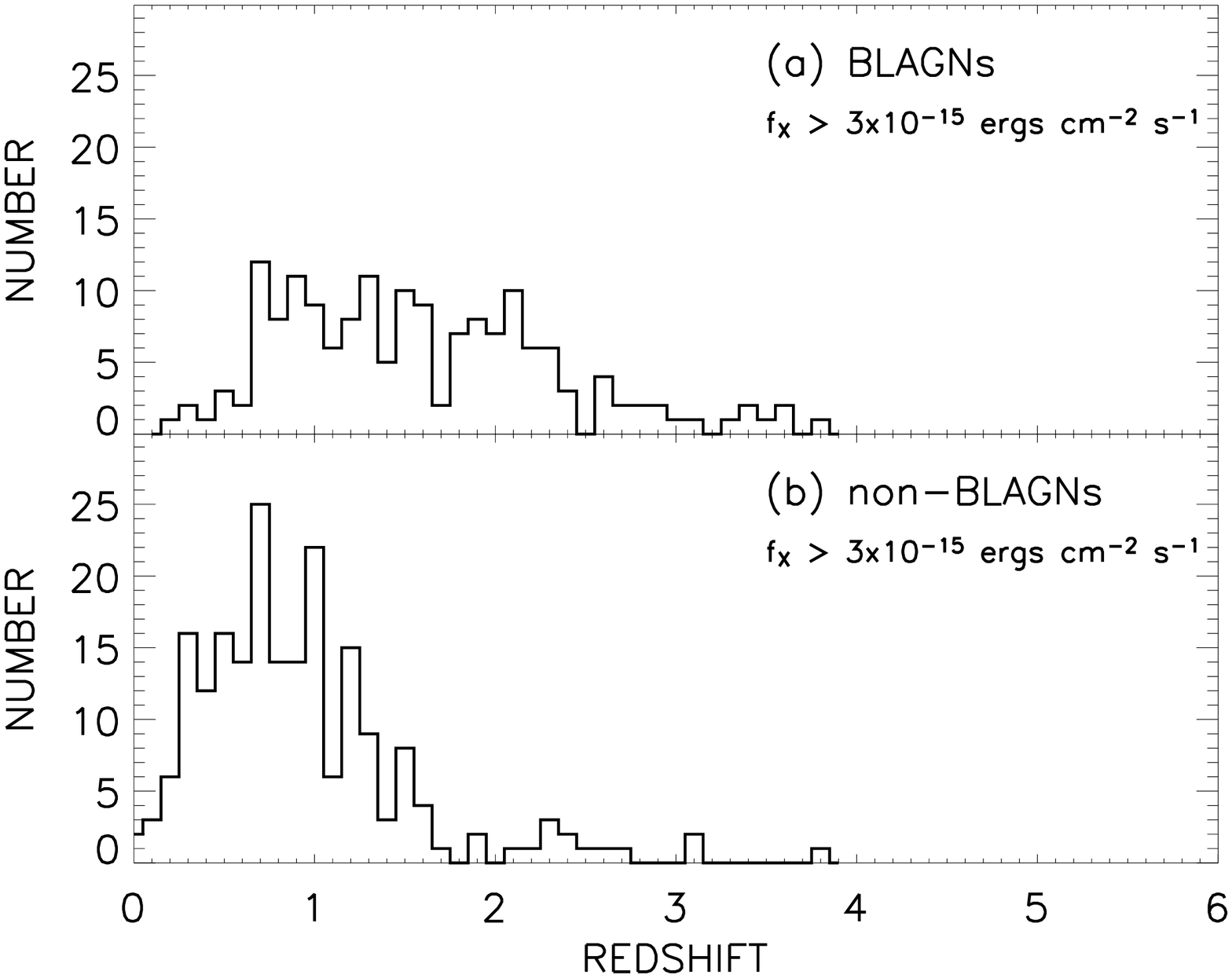}
  \caption{\label{redshift_hist_type}Redshift distribution of
    ({\it a}) BLAGNs and ({\it b}) non-BLAGNs with $f_{2 - 8~{\rm keV}} > 3
    \times 10^{-15}\ \flux$ (see \S~6) in four surveys: CLASXS, 
    CDF-N (Barger et al.\ 2003), CDF-S (Szokoly et al.\ 2004), 
    and {\it XMM-Newton} Lockman Hole (Mainieri et al.\ 2002).  
    All stars and sources without measured redshifts have been excluded.}
\end{figure}

Figure~\ref{r_mag_vs_redshift} shows redshift versus $R$-band
magnitude for the CLASXS hard \mbox{X-ray} sample, with BLAGNs denoted
by solid squares and non-BLAGNs denoted by open squares.
Figure~\ref{redshift_hist_type} combines the CLASXS data with the data
of \citet{barger03}, \citet{mainieri02}, and \citet{szokoly04},
dividing the \mbox{X-ray} sources into BLAGNs and non-BLAGNs.  To
remove the bias introduced by the depth of the CDF-N survey, which is
deep enough to detect the \mbox{X-ray} emission from nearby galaxies
with no AGN component, we have only shown the sources with
$f_{2-8~{\rm keV}} > 3 \times 10^{-15}$~\flux.  The combined surveys
have very few BLAGNs at $z<0.5$, but the numbers increase at $z=0.5$
and appear to decrease only slightly over the range $z\sim0.5-2.5$ before
falling off beyond.  The redshift distribution is quite different for
the non-BLAGN population.  However, there are strong selection effects
in the spectroscopic identifications.  There are many more AGNs at low
redshifts that do not have broad optical emission lines, either
because of low intrinsic AGN luminosity or obscuration.  Thus, the identifications
for these sources are based on their galaxy spectra, and at high
redshifts, galaxy spectra are much harder to spectroscopically
identify than BLAGN spectra, particularly in the spectroscopic
redshift desert at $z=1.5-2$ \citep[e.g.,][]{wirth04,cowie04b}.

%
%

\section{Colors and Luminosities of the CLASXS sources}

%
%
\begin{figure*}[th]
  \epsscale{1.0}
   \includegraphics[angle=90,scale=0.6]{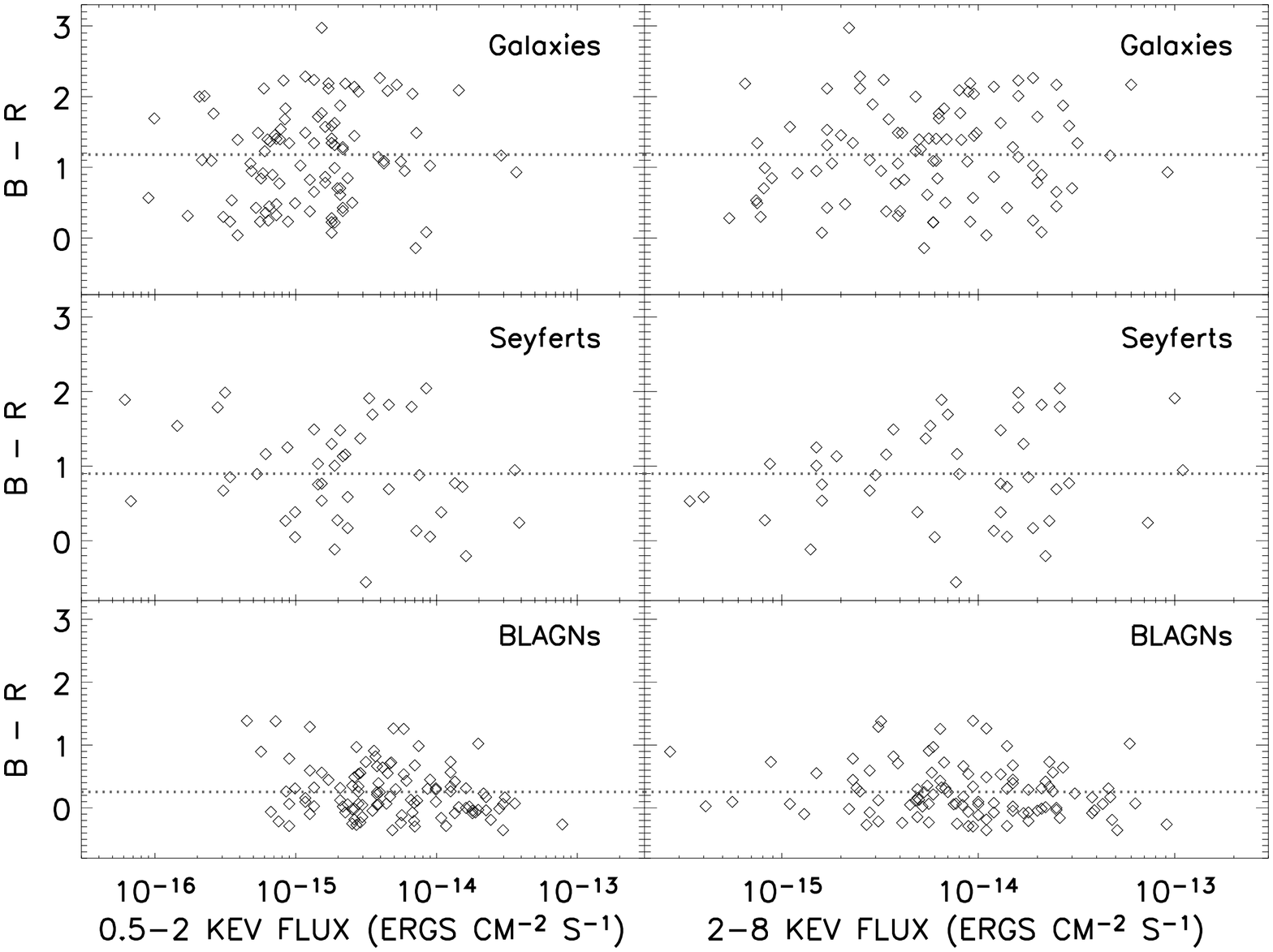}
  \caption{\label{x-ray_flux_vs_B-R}\emph{B$-$R} color vs. soft {\em (left)}
    and hard {\em (right)} \mbox{X-ray} flux for three spectral classes (normal
    galaxies, Seyferts, and BLAGNs) for the CLASXS sources.  The
    dotted lines denote the median $B-R$ colors for each spectral class.}
\end{figure*}

In Figures~\ref{x-ray_flux_vs_B-R}{\it a} and
\ref{x-ray_flux_vs_B-R}{\it b}, we plot the observed {\it B$-$R} color
versus soft and hard flux for the spectroscopically identified CLASXS
sources with $R < 26$, excluding any sources that are saturated.  We
break the sources into three spectroscopic categories: BLAGNs,
Seyfert galaxies, and ``normal'' galaxies (the last category combines the
absorbers and star formers classes of \S~6).  The BLAGNs are, on average, much
bluer than the other types of sources.  This is due to the excess blue
continuum from the AGN itself that is commonly observed in unobscured
(type I) AGNs.  Thus, the colors and the spectroscopic classes are in
good agreement.

%
%
\begin{figure}[t]
  \epsscale{1.0}
   \includegraphics[angle=90,scale=0.6]{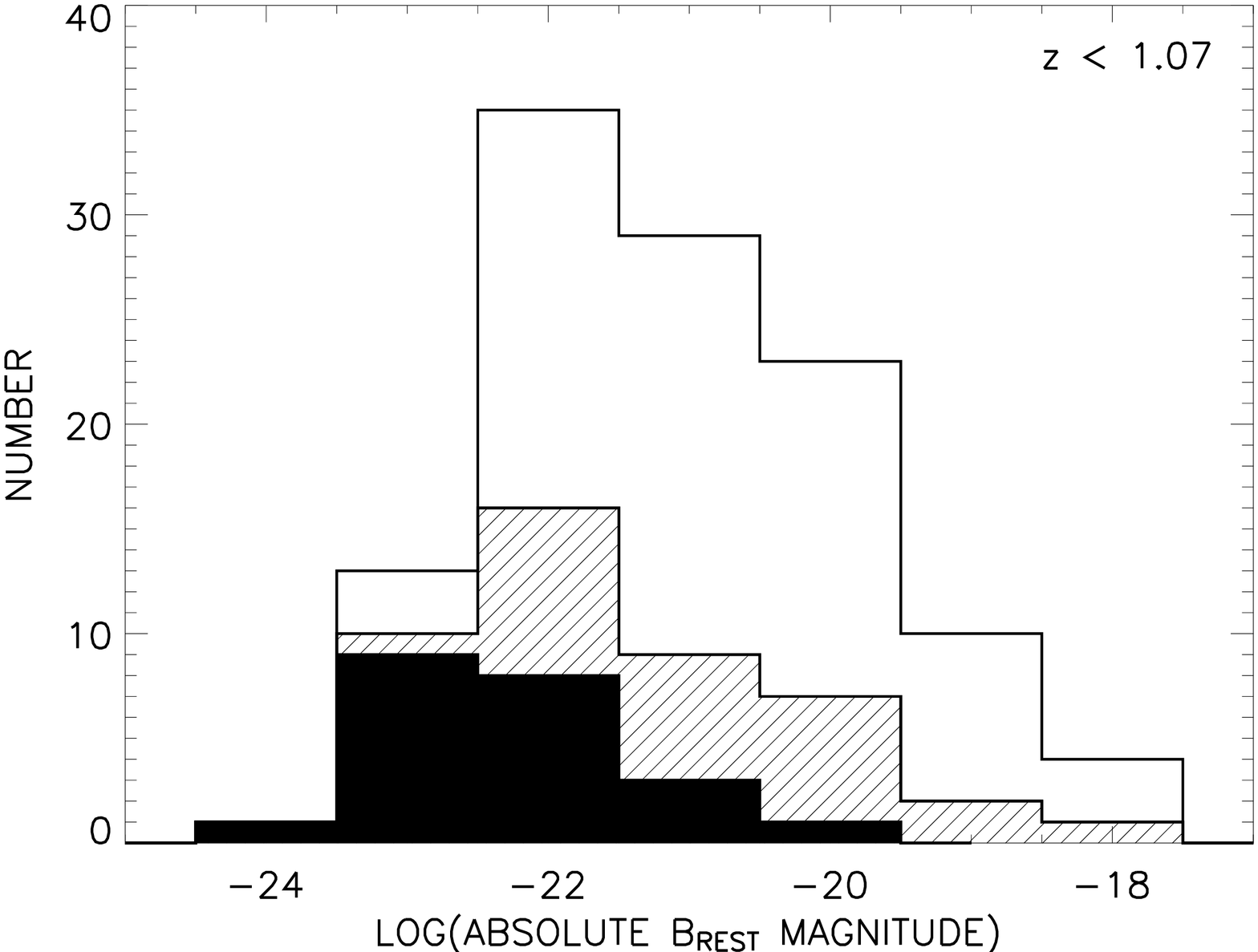}
  \caption{\label{abs_b_dist}Distribution of absolute rest-frame
    $B$ magnitude for all of the CLASXS sources with measured
    redshifts less than $z=1.07$ {\em (open histogram)\/}.  BLAGNs
    (non-BLAGNs) are shown as a solid ({\it hatched}) histogram.  (See \S~7
    for an explanation of the redshift limit.)}
\end{figure}

The distribution of the absolute rest-frame \emph{B} magnitudes for
the CLASXS sources is shown in Figure~\ref{abs_b_dist}, calculated
using the measured redshift and interpolating the \emph{B}-magnitude
in the observer's rest-frame from the observed broadband colors.
Since we do not have near-infrared observations of all of our
\mbox{X-ray} sources, we can only use this technique to measure
absolute \emph{B}-magnitudes for sources with $z<1.07$.  Beyond this
redshift, the rest-frame \emph{B}-band moves redward beyond our $z'$
observations.  Only non-stellar sources with measured magnitudes in at
least two wavebands are included.  BLAGNs and non-BLAGNs are shown as
solid and hatched histograms, respectively.

Our photometric data do not have the resolution necessary to separate
the luminosity of the host galaxy from that of the central AGN, so we
can only comment on the combined AGN+host luminosity.  Since BLAGNs
have an excess blue continuum, one would expect a larger number of
BLAGNs at brighter absolute \emph{B} magnitudes.  Indeed,
Figure~\ref{abs_b_dist} shows that this is the case.

\section{Discussion}

%
%
\begin{figure*}[th]
  \epsscale{0.7}
  \plotone{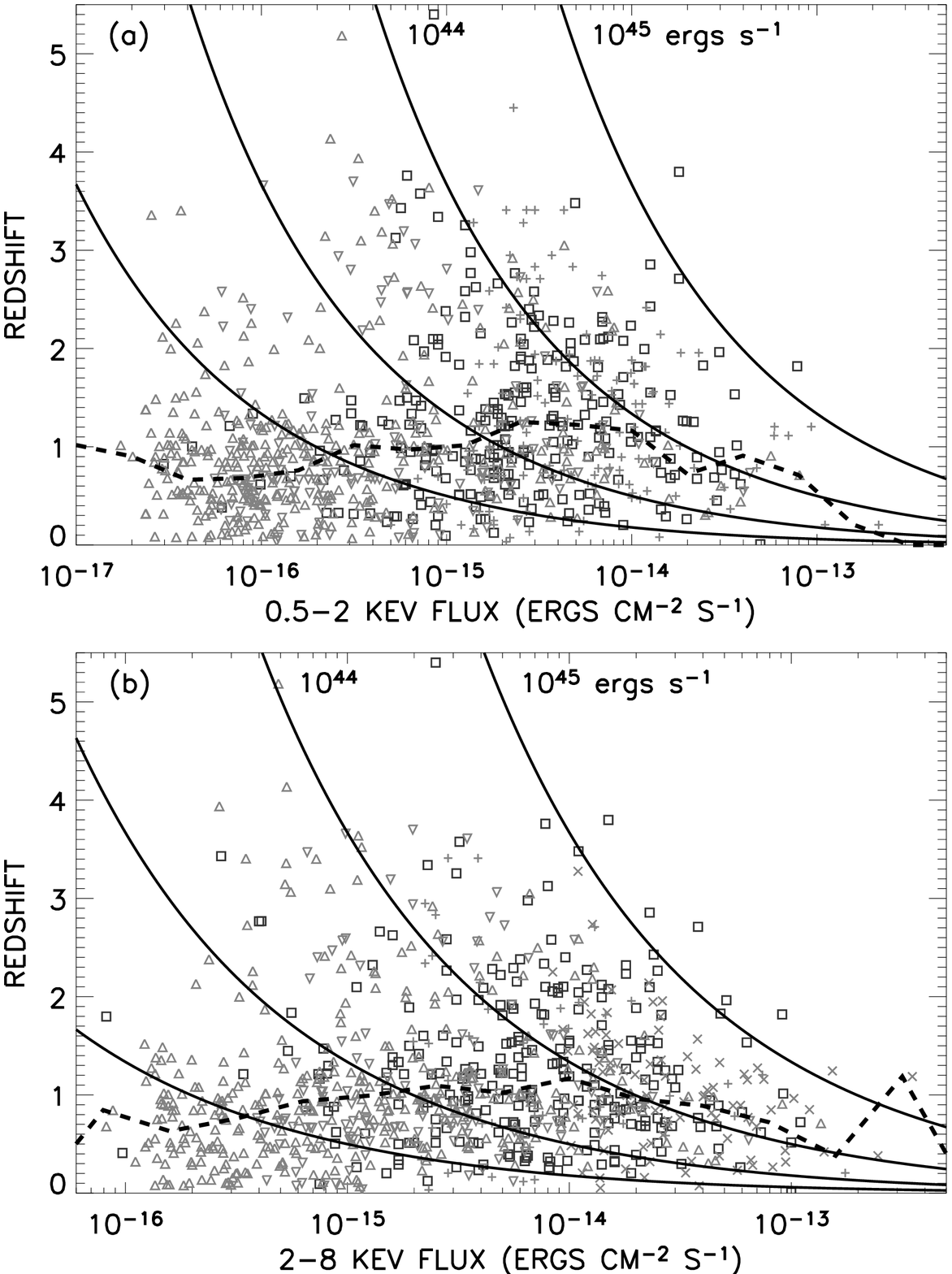}
  \caption{\label{redshift_flux}({\it a}) Redshift vs. $0.5 - 2.0$~keV flux
    for the soft \mbox{X-ray}--selected sources in CLASXS ({\it
      squares}), the CDF-N ({\it upward-pointing triangles}; Barger et al.
    2003), the CDF-S ({\it downward-pointing triangles}; Szokoly et al. 2004),
    and in combined {\it ROSAT} (Lehmann et al. 2001), {\it
      XMM-Newton} (Mainieri et al. 2002), and {\it
      ASCA} (Akiyama et al. 2003) surveys ({\it plus signs}).  ({\it b})
    Redshift vs. $2-8$~keV flux for the hard \mbox{X-ray} selected
    sources (symbols as in (a); Lehmann et al. {\it ROSAT} sources
    have no hard \mbox{X-ray} measurements).  In
    ({\it b}) we added the hard \mbox{X-ray} data from the HELLAS2XMM survey
    ({\it crosses}; Fiore et al.\ 2003).  Dashed lines show
    the median redshift of the combined samples versus soft and hard
    \mbox{X-ray} flux. [{\it A color version of this figure is
    available in the electronic edition of the Astronomical Journal.}]}
\end{figure*}

%
%
\begin{figure*}[hbp]
\epsscale{1.0}
   \includegraphics[angle=90,scale=0.6]{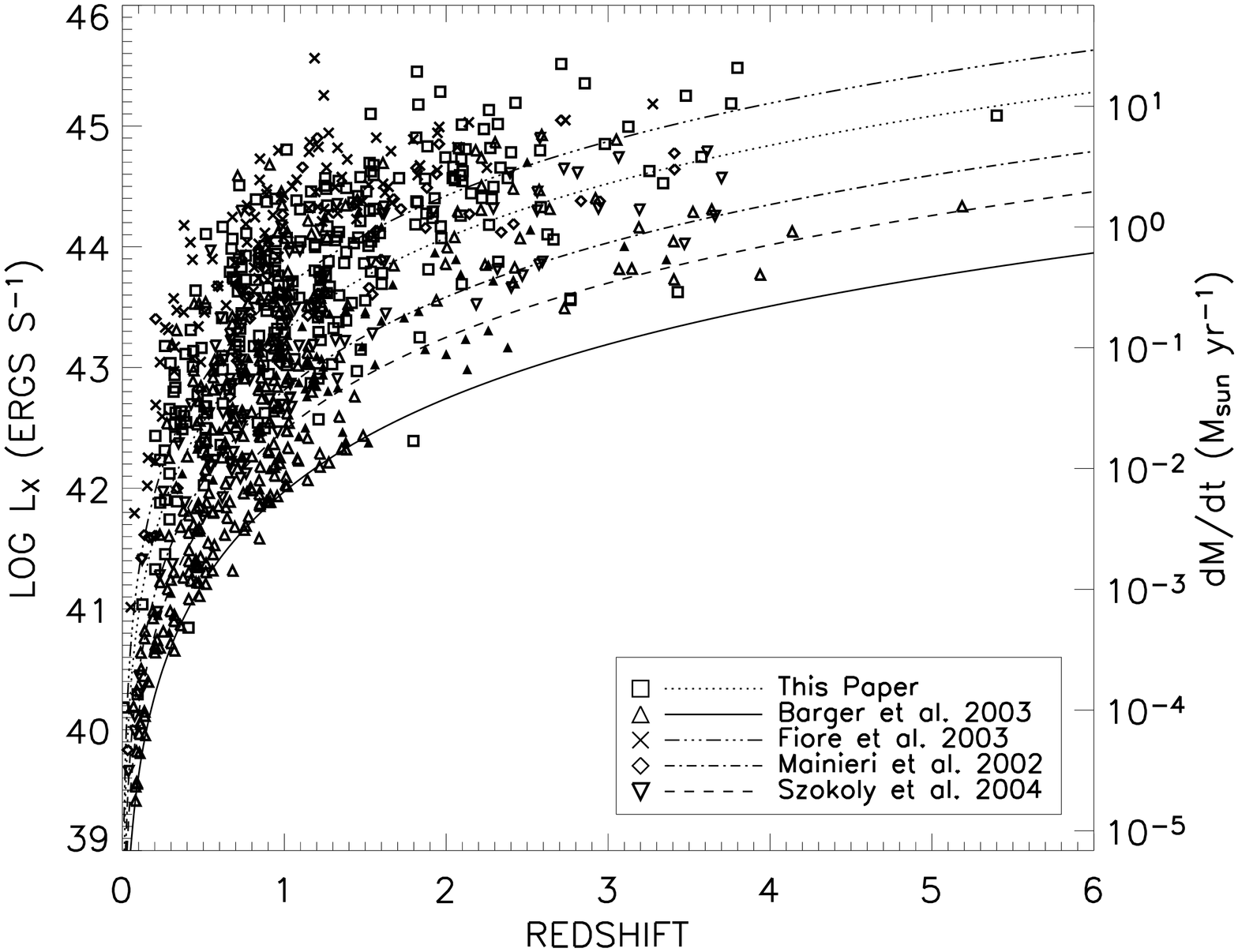}
\caption{\label{lum_z} Rest-frame hard \mbox{X-ray} luminosity of
  the CLASXS {(\it squares)}, CDF-N \citep[{\it upward-pointing
    triangles};][]{barger03}, HELLAS2XMM \citep[{\it
    crosses};][]{fiore03}, {\it XMM-Newton} Lockman Hole \citep[{\it
    diamonds};][]{mainieri02}, and CDF-S \citep[{\it downward-pointing
    triangles};][]{szokoly04} sources with known redshifts.  CDF-N
  sources with spectroscopic (photometric) redshifts are shown as open
  ({\it filled}) symbols.  Curves represent the on-axis flux limits for the
  different surveys (see legend).  Right-hand axis shows the accretion
  rate (in solar masses per year).  [{\it A color version of this figure is
    available in the electronic edition of the Astronomical Journal.}]}
\end{figure*}

%
%

In Figures~\ref{redshift_flux}{\it a} and \ref{redshift_flux}{\it b}, we
plot redshift versus soft and hard \mbox{X-ray} flux for the combined
samples of \S~5. The data and symbols are identical to those in
Figure~\ref{r_mag_vs_sb}.  The CLASXS data nicely fill in the
intermediate flux interval between the deep \emph{Chandra} data and
the shallower \emph{ASCA} and \emph{ROSAT} data.  The dashed lines
show the median redshifts of the combined samples versus soft and hard
\mbox{X-ray} flux.  The median redshifts remain around $z\sim1$ and do
not appear to be strongly correlated with either soft or hard
\mbox{X-ray} flux.  Of course, the redshift determinations depend on
the optical magnitudes of the sources.  Since the magnitudes of
high-redshift sources may be faint, these surveys might be biased
against finding high-redshift sources.  The solid lines denote
constant rest-frame \mbox{X-ray} luminosities, which were calculated
assuming a $\Gamma = 1.8$ power-law spectral energy distribution with
no absorption.  They range from $10^{42}$ to $10^{45}$~ergs~s$^{-1}$;
the two largest luminosities are labeled.  Any source more luminous
than $L_X=10^{42}$~ergs~s$^{-1}$ is very likely to be an AGN on
energetic grounds \citep{zezas98,moran99}, though many of the
intermediate luminosity sources do not show obvious AGN signatures in
their optical spectra \citep[e.g.,][]{barger01a,horn01,tozzi01}.

We use the hard \mbox{X-ray} fluxes and redshifts of the combined
samples in Figure~\ref{redshift_flux}{\it b} to calculate the
rest-frame hard \mbox{X-ray} luminosities.  As in \S~2, we assume a
power law spectral energy distribution with $\Gamma = 1.8$ for
calculating the {\it K}-corrections to determine the rest-frame hard
\mbox{X-ray} luminosities.  Using a universal power law index rather
than individual indices to calculate the {\it K}-corrections results
in only a small difference in the calculated luminosities 
\citep{barger02}.  Figure~\ref{lum_z} shows the rest-frame hard
\mbox{X-ray} luminosities for the hard \mbox{X-ray} sources and
limiting fluxes of the CLASXS, \citet{barger03}, \citet{fiore03},
\citet{mainieri02}, and \citet{szokoly04} catalogs.  The CLASXS
\mbox{X-ray} sample has the largest number of $L_{2-8~{\rm keV}} \ge
10^{45}~{\rm ergs~s}^{-1}$ sources.  Even with the increased solid
angle covered by the five surveys, there is an obvious lack of
extremely luminous ($L_{2-8~{\rm keV}} \ge 10^{44}~{\rm ergs~s}^{-1}$)
AGNs at $z<0.5$.

The bolometric luminosity, $L_{BOL}$, of an AGN is related to the rate
of accretion, $\dot{M}$, onto its supermassive black hole using the
expression $\epsilon \dot{M}_{BH} = L_{BOL}/c^{2}$, where $c$ is the
speed of light and $\epsilon$ is the accretion efficiency (typically,
$\epsilon \sim 0.1$).  Because AGNs are luminous from radio to
\mbox{X-ray} energies, multiwavelength observations are needed to
accurately calculate their bolometric luminosities. If we use a
bolometric correction of 35 \citep{elvis94,barger01a,kuraszkiewicz03}
appropriate for the type I AGNs to translate hard \mbox{X-ray}
luminosity, $L_{X}$, to $L_{BOL}$ and assume an efficiency of
$\epsilon =0.1$, then we can roughly estimate the accretion rates.
The results are shown in Figure~11 {\em (right-hand axis)\/}.  The
small solid angle of the CDF-N survey prevented \citet{barger01b} from
determining if the lack of low redshift, high luminosity sources in
their sample were due to an actual deficiency of these sources at low
redshifts or simply a selection effect owing to the small local volume
sampled and the low spatial density of these sources.  CLASXS samples a
larger local cosmological volume to better constrain the accretion
rate at low redshifts, and we can now clearly see that the deficiency
of these sources is indeed real and that at $z<0.5$ there are very few
sources with accretion rates above a solar mass per year.

\section{Summary}

We have presented optical photometric and spectroscopic observations
of the \mbox{X-ray} sources in the CLASXS survey.  We obtained
$BVRIz'$ photometry for 521 of the 525 \mbox{X-ray} sources in
the sample.  We spectroscopically observed 467 of the CLASXS sources,
obtaining redshift identifications for 271.  The positions of the
sources in our optical catalog agree well with the \mbox{X-ray}
catalog, with average astrometric offsets of $0.0 \pm 0.5$ arcseconds
in both right ascension and declination.

Using our optical data, we have examined the relationship between the
\mbox{X-ray} and optical source properties.  We find that the average
soft \mbox{X-ray}--to--optical flux ratio remains relatively constant
for CLASXS sources with low to moderate column densities of obscuring
material.  Conversely, the hard \mbox{X-ray}--to--optical flux ratio
for these sources appears to be correlated with line-of-sight column
density.  Because of the large solid angle covered by CLASXS, we do
not find the low-redshift structures that are seen in narrower, deeper
{\it Chandra} surveys.  However, we do detect a possible excess of
sources at $z=1.8$.  Using hard \mbox{X-ray} catalogs from five {\it
  Chandra} and {\it XMM-Newton} surveys we find that there is a
distinct lack of luminous, high-accretion rate sources at low
redshifts ($z<1)$, consistent with previous observations which showed
that supermassive black hole growth is dominated at low redshifts by
sources with low accretion rates \citep{barger01b}.

\acknowledgements
We thank the referee, David Alexander, for helpful comments that
improved the manuscript.  We gratefully acknowledge support from
NASA's National Space Grant College and Fellowship Program and the
Wisconsin Space Grant Consortium (A.~T.~S.), CXC grants GO2-3191A
(A.~J.~B.) and GO2-3187B (L.~L.~C.), NSF grants AST-0084847 
and AST-0239425 (A.~J.~B.) and AST-0084816 (L.~L.~C.),
the University of Wisconsin Research Committee with funds granted by
the Wisconsin Alumni Research Foundation (A.~J.~B.), the Alfred P.
Sloan Foundation (A.~J.~B.), the David and Lucile Packard Foundation
(A.~J.~B.), and the IDS program of R.~F.~M.

\appendix
\section{Appendix}
\label{appendix_A}

Details of the Lockman Hole-Northwest photometric observations and
reductions can be found in \citet{capak04b}.  We
provide a short summary of the Subaru and CFHT broadband photometric
observations here.  For the Subaru images, bad pixels and chip gaps
were filled in by creating mosaics using a five point star-shaped
dither pattern with $1\arcmin$ steps.  The separation between
dithered images was made large enough to create deep sky flats during
the image reduction process.  The camera was rotated by $90\deg$
between dithers to remove bleeding from bright stars and provide
better photometric calibration.  The Suprime-Cam data cover the
central $36^{\prime} \times 36^{\prime}$ of the field, but only the
central $27^{\prime} \times 27^{\prime}$ of the Suprime-Cam field goes
to the full depth due to vignetting in Suprime-Cam.  In the
\emph{R} band, we added four Suprime-Cam pointings covering
$54^{\prime} \times 54^{\prime}$ from the Subaru archive.  These data
were collected using the same dither pattern and reduced by us in the
same manner as our other data.  For the CFHT observations, a standard
seven point circular dither pattern was used.  The radius of the
dither circle was $7.5^{\prime\prime}$, which was enough to fill the
chip gaps but minimized the chip overlap.  The observing time was
divided between two pointings separated by $14^{\prime}$ in the
north-south direction.  This allowed us to cover a $42^{\prime} \times
42^{\prime}$ field while going deeper in a central strip.

All the data were reduced using Nick Kaiser's IMCAT tools.
Flat-fielding of the CFHT images was done using
ELIXIR\footnote{Available at http://www.cfht.hawaii.edu/Instruments/Elixir/}.  The
$z'$ data were corrected for fringing.  The images were aligned
astrometrically using a self-consistent astrometric solution described
in \citet{capak04a}.  The solution was calculated using the CFHT
$R$-band data along with wide-field Suprime-Cam data. The USNO-B1.0
\citep{monet03} was used as the reference catalog for this process and
provided the absolute astrometric zeropoint. All other data were
warped onto this $R$-band image. The individual mosaics were then
combined.  Thumbnails of 521 sources are shown in Figure~\ref{thumbnails}.

%
%
\begin{figure*}[ht]
  \epsscale{0.8}
  \plotone{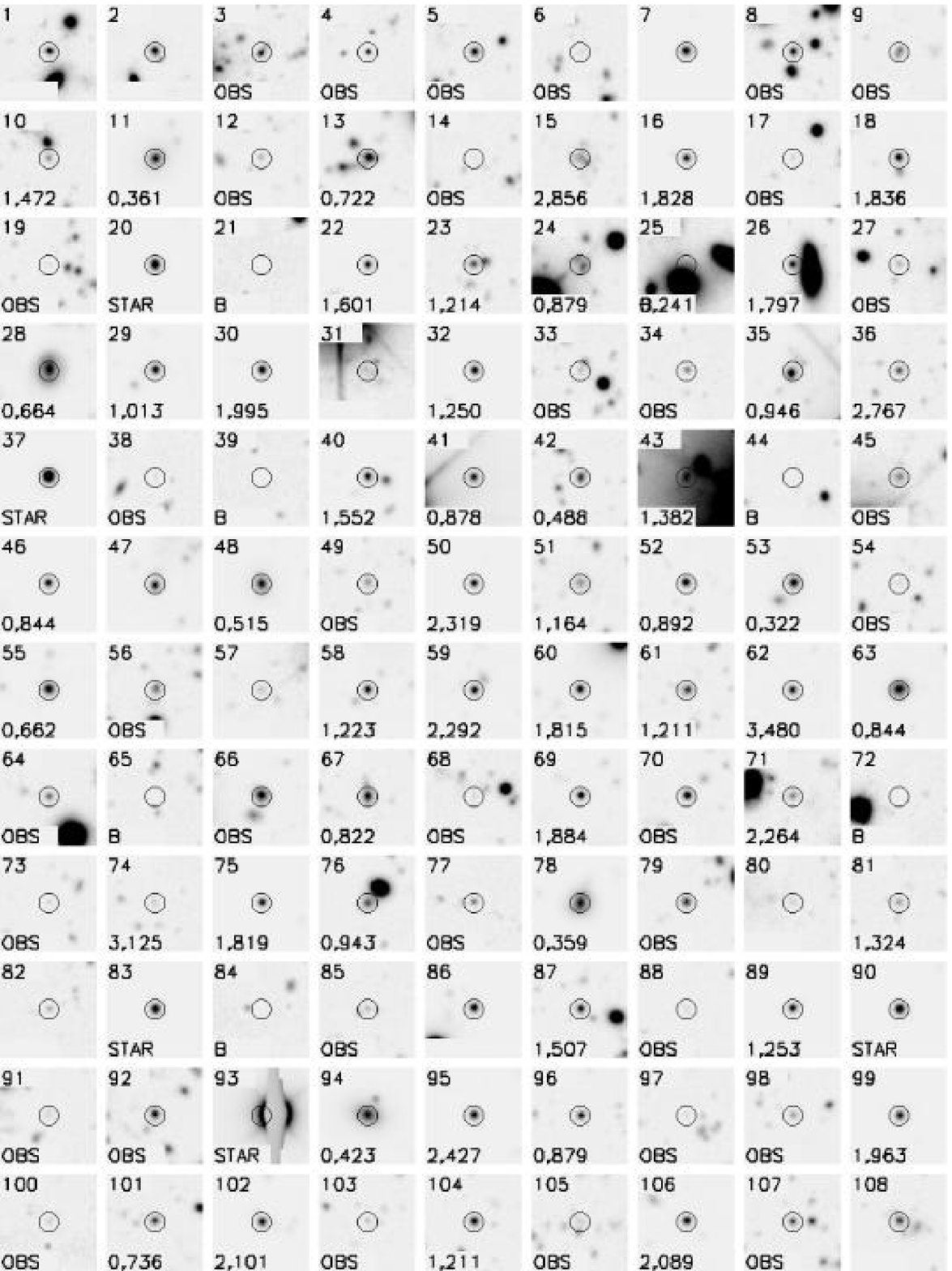}
  \caption{\label{thumbnails}$R$-band thumbnails of 521 of the 525
    CLASXS.  Four \mbox{X-ray} sources (192, 318, 328, and 359) are
    outside the fields-of-view of our optical images.  The $R$-band $2
    \sigma$ AB magnitude limit is 27.9.  Thumbnails are 15\arcsec\ on
    a side.  Y04 source number (column [1] of Table~\ref{main_table}
    in this paper) is displayed in the top left corner.  If an
    \mbox{X-ray} source has an optical counterpart, then a 3\arcsec\ 
    aperture is shown centered on the optical counterpart.  If there
    is no detected counterpart, then the 3\arcsec\ aperture is
    centered on the \mbox{X-ray} position.  Redshifts for identified
    optical counterparts are shown in the bottom left corner.  Stars
    are labeled ``star''. Optically undetected sources are labeled
    ``B'' for blank.  Spectroscopically observed sources that could
    not be identified are labeled ``obs''.}
\end{figure*}

\begin{figure*}[ht]
  \epsscale{0.8}
  \figurenum{\ref{thumbnails}}
  \plotone{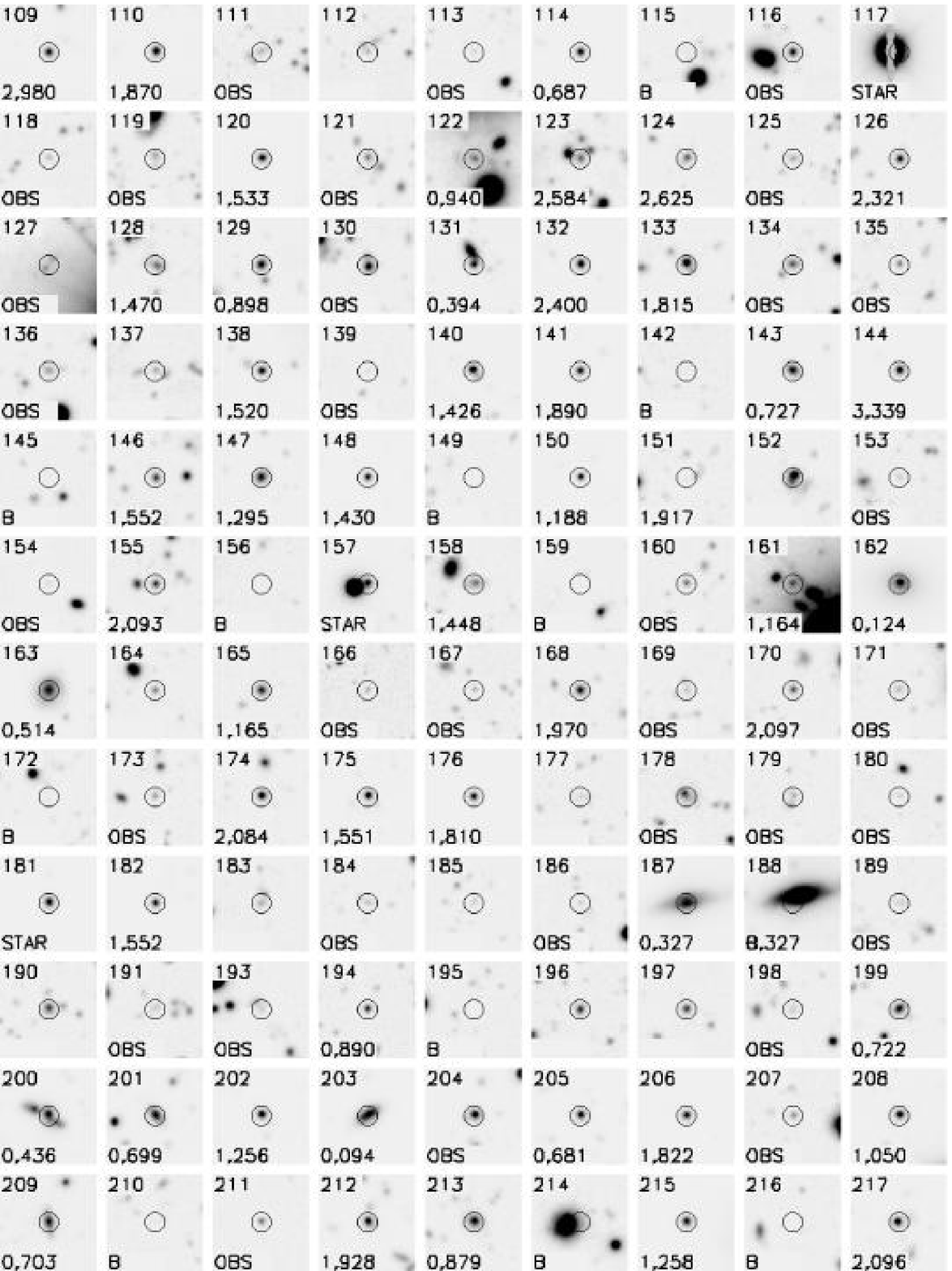}
  \caption{cont'd}
\end{figure*}

\begin{figure*}[ht]
  \epsscale{0.8}
  \figurenum{\ref{thumbnails}}
  \plotone{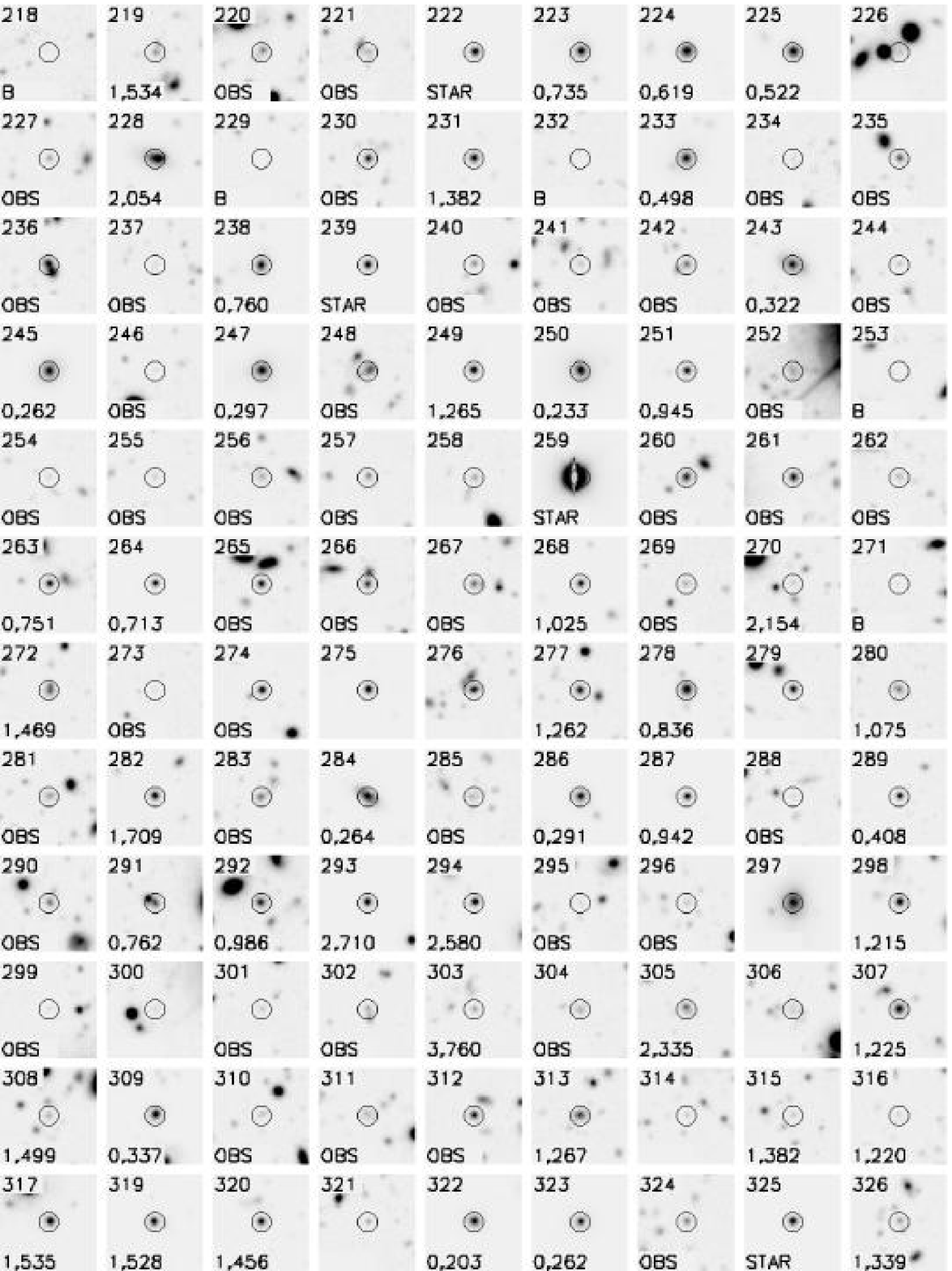}
  \caption{cont'd}
\end{figure*}

\begin{figure*}[htp]
  \epsscale{0.8}
  \figurenum{\ref{thumbnails}}
  \plotone{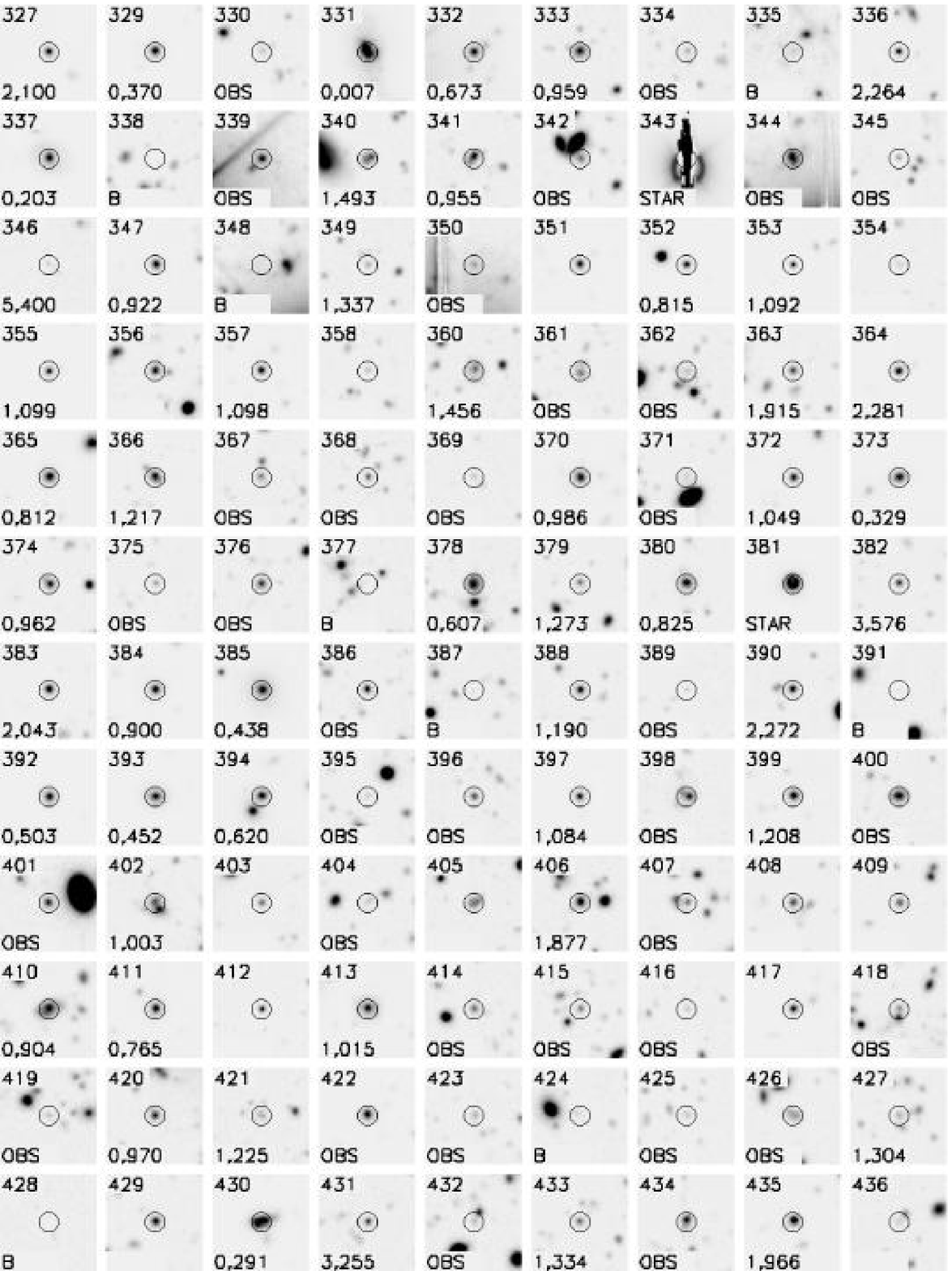}
  \caption{cont'd}
\end{figure*}
 
\begin{figure*}[htp]
  \epsscale{0.8}
  \figurenum{\ref{thumbnails}}
  \plotone{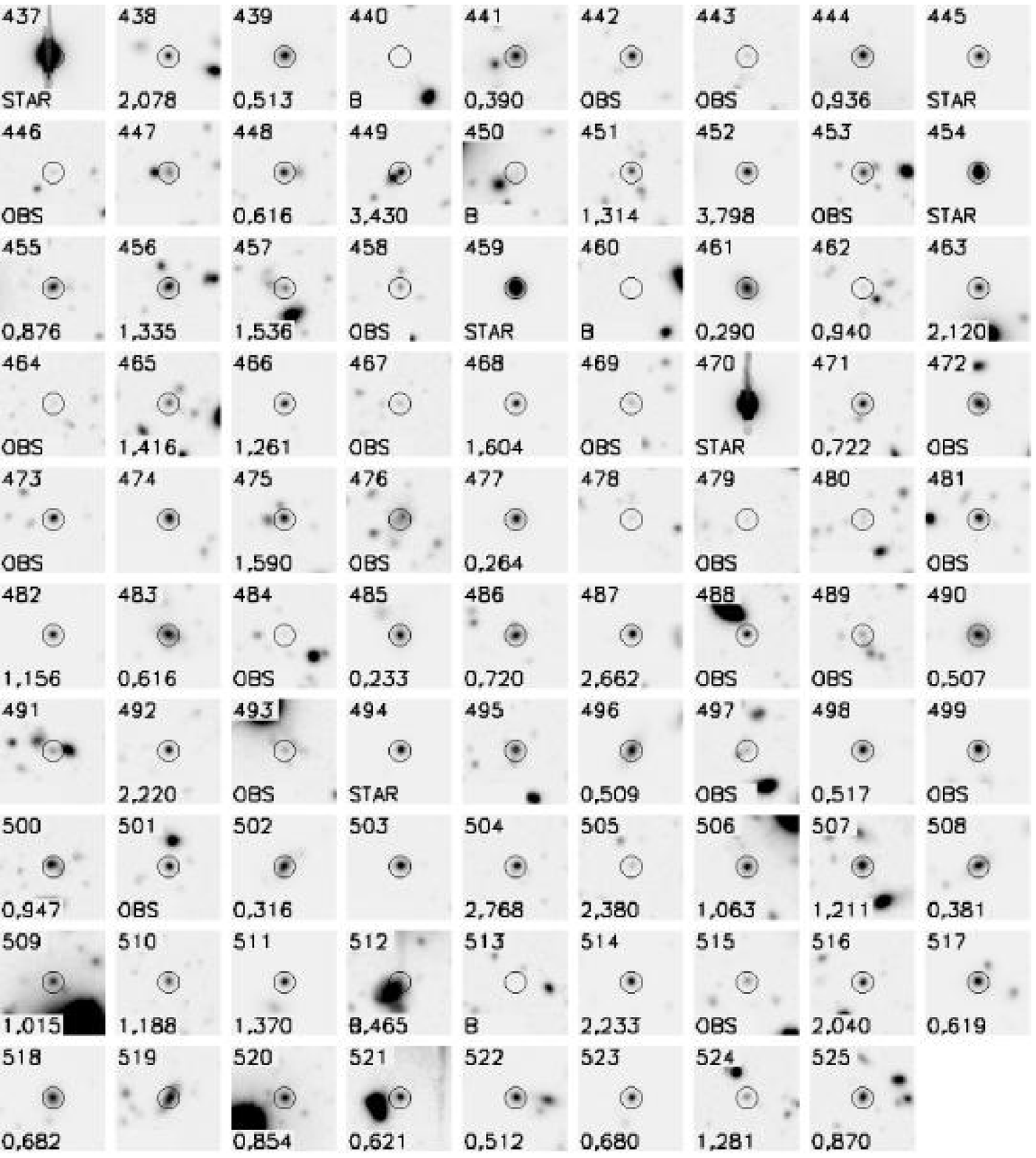}
  \caption{cont'd}
\end{figure*}

In Table~\ref{main_table} we present optical magnitudes and
spectroscopic measurements, where available, for the CLASXS
\mbox{X-ray} point source catalog of Y04. The 525 \mbox{X-ray} sources
are ordered by increasing right ascension and are labeled with the Y04
source number (col.~[1]).  The right ascension and declination
coordinates of the \mbox{X-ray} sources are given in columns
(2)~and~(3), and the smallest off-axis angle is given in column (4).
The right ascension and declination coordinates of the optical
counterparts are given in decimal degrees in columns (5)~and~(6).  The
soft ($0.5-2$~keV) and hard ($2-8$~keV) \mbox{X-ray} fluxes,
abbreviated SB and HB, are given in columns~(7)~and~(8).  The aperture
corrected, broadband \emph{B, V, R, I, {\rm and} $z'$} apparent
magnitudes are given in columns~(9)-(13).  The aperture correction was calculated by
examining the curve-of-growth for isolated, moderately bright ($R =
20-26$) sources.  The 3\arcsec\ aperture corrections for the Subaru
(CFHT) images are $-0.17$ ($-0.30$), $-0.22$, $-0.11$ ($-0.15$),
$-0.33$, $-0.13$ ($-0.22$), for the \emph{B}, \emph{V}, \emph{R},
\emph{I}, and $z'$ bands, respectively.  Sources that fall outside of
the field-of-view of a given filter are left blank.  If a source falls
within the field-of-view, but is not detected, then it is assigned the
$2\sigma$ magnitude limit for that filter and is considered an upper
limit.  The spectroscopic redshifts are given in column~(14).

\vfil\eject\clearpage


%
\begin{deluxetable}{lcccccrrrrrrrc}
\tablecaption{\label{main_table}Optical Properties of the CLASXS \mbox{X-ray} Sample}
\tabletypesize{\scriptsize}
\rotate
\tablewidth{0pt}
\tablehead{\colhead{} &
           \multicolumn{2}{c}{X-ray (J2000.0)} &
           \colhead{} &
           \multicolumn{2}{c}{Optical (J2000.0)} &
           \colhead{} &
           \colhead{} &
           \colhead{} &
           \colhead{} &
           \colhead{} &
           \colhead{} &
           \colhead{} &
           \colhead{} \\
           \colhead{Y04} &
           \colhead{R.A.} &
           \colhead{Decl.} &
           \colhead{$R_{aim}$\tablenotemark{a}} &
           \colhead{R.A.} &
           \colhead{Decl.}  &
           \colhead{\emph{SB}\,\tablenotemark{b}} &
           \colhead{\emph{HB}\,\tablenotemark{c}} &
           \colhead{\emph{B}} &
           \colhead{\emph{V}} &
            \colhead{\emph{R}} &
           \colhead{\emph{I}} &
           \colhead{$z'$} &
           \colhead{$z_{spec}$\,\tablenotemark{d}}  \\
           \colhead{(1)} &
           \colhead{(2)} &
           \colhead{(3)} &
           \colhead{(4)} &
           \colhead{(5)} &
           \colhead{(6)} &
           \colhead{(7)} &
           \colhead{(8)} &
           \colhead{(9)} &
           \colhead{(10)} &
           \colhead{(11)} &
           \colhead{(12)} &
           \colhead{(13)} &
           \colhead{(14)}}  
\startdata
  1 & 157.731750 & 57.55556 & 9.99 & 157.730803 & 57.55546 &      3.42 &      3.70 & \nodata & \nodata &   23.1 & \nodata & \nodata & \nodata \\
  2 & 157.748667 & 57.64561 & 8.76 & 157.747566 & 57.64558 &      1.62 &      5.30 & \nodata & \nodata &   22.7 & \nodata & \nodata & \nodata \\
  3 & 157.763833 & 57.61400 & 8.16 & 157.763506 & 57.61356 &      2.88 &     13.00 & \nodata & \nodata &   23.5 & \nodata & \nodata & obs \\
  4 & 157.775250 & 57.63011 & 7.80 & 157.774430 & 57.63017 &      3.33 &      0.74 & \nodata & \nodata &   23.8 & \nodata &   23.4 & obs \\
  5 & 157.799292 & 57.58936 & 7.27 & 157.799617 & 57.58951 &      2.88 &      0.91 &   24.7 & \nodata &   24.0 & \nodata &   23.0 & obs \\
\enddata
\tablecomments{The complete version of this table is in the electronic
edition of the Astronomical Journal.  The printed edition contains only a sample.}
\tablenotetext{a}{Radial distance from aim point in arcminutes}
\tablenotetext{b}{Soft Band (SB) X-ray flux in units of $10^{-15}$ \flux}
\tablenotetext{c}{Hard Band (HB) X-ray flux in units of $10^{-15}$ \flux}
\tablenotetext{d}{Spectroscopically identified stars are labeled
  ``star''.  If the redshift for a spectroscopically observed source
  could not bo determined from the spectrum, then the source is
  labeled ``obs''.}
\tablenotetext{e}{Source 151: This source has a measured redshift but
  no optical counterpart was detected by SExtractor in our deep
  $R$-band image.  Upon closer examination, we find a very faint
  optical source 0.5\arcsec\ to the North of the \mbox{X-ray} position.  We
  believe SExtractor failed to detect this faint source because it
  blended with the brighter source to the Northwest during the
  Gaussian smoothing that was performed on the image pre-detection.
  We conclude that this very faint optical source is the counterpart
  to the \mbox{X-ray} source and include its optical position in the
  catalog.}
\tablenotetext{f}{Sources 187 and 188: These two sources appear to be
  located in the same host galaxy.  Source 187 is the \mbox{X-ray}
  source in the center of the galaxy, while source 188 appears to be
  an off-nuclear source.  These sources appear distinct in our
  \mbox{X-ray} images, and we conclude that this is not a double
  detection of a single \mbox{X-ray} source.  We are uncertain if
  source 188 is associated with the galaxy or if it is a foreground or
  background object.  Because of this uncertainty, we do not assign
  the measured redshift for the galaxy to source 188.}
\tablenotetext{g}{Source 413: There are two separate redshift systems
  for this source.  The first system is at z=0.0077 and has strong
  H$\alpha$ and weak [NII] and [SII].  This would seem to be the
  bright galaxy in the thumbnail.  The second system is at z=1.0152 has
  very strong [OIII], though much of the Balmer series can also be
  seen. [OII] is not visible.  This object is slightly spatially
  displaced in the spectral image from the H$\alpha$ system.  We have
  chosen to use the high redshift system in the table, since the spectrum
  seems more plausible for the \mbox{X-ray} source, but this could be
  a misidentification, or lensing could be a problem.  This source
  should be treated as suspect.}

\end{deluxetable}

\vfil\eject\clearpage

%
%
\begin{deluxetable}{lllr}
\tablewidth{0pt}
\tablecaption{\label{data_details1}Summary of Optical Imaging
  Observations: Telescopes, Dates of Observations, and Integration Times}
\tablehead{
\colhead{Band} &
\colhead{Telescope} &
\colhead{UT Dates}&
\colhead{Integration}\\
\colhead{} &
\colhead{}&
\colhead{} &
\colhead{(hr)}}

\startdata
$B$          & Subaru 8.2~m & 2001 February 27-28 & 1.7  \\
$B$          & CFHT 3.6~m   & 2002 December 4-5,~27-29 & 5.8 \\
$V$          & Subaru 8.2~m & 2002 April 22-23 & 6.4  \\
$R$          & Subaru 8.2~m & 2001 February 27-28 & 5.2  \\
$R$          & Subaru 8.2~m & 2001 March 19-20 & 2.0  \\
$R$          & CFHT 3.6~m   & 2003 January 2-3,~25-30 & 11.9 \\
$I$          & Subaru 8.2~m & 2001 April 21-22 & 0.9  \\
$z^{\prime}$ & Subaru 8.2~m & 2001 April 21, 23 & 1.3  \\
$Z$          & CFHT 3.6~m   & 2002 December 27 & 11.9 \\
$Z$          & CFHT 3.6~m   & 2003 January 25-31 & 11.9 
\enddata
\end{deluxetable}

%
%
\begin{deluxetable}{llcccc}
\tablewidth{0pt}
\tablecaption{\label{data_details2}Summary of Optical Images: Filters,
Telescopes, Data Quality, Depth, and Area Coverage}
\tablehead{
\colhead{Band} &
\colhead{Telescope} &
\colhead{Average seeing} &
\colhead{$2~\sigma$ limit} &
\colhead{Total area} &
\colhead{Deep area} \\
\colhead{} &
\colhead{} &
\colhead{(arcsecond)} &
\colhead{(AB mag)} &
\colhead{(deg$^{2}$)} &
\colhead{(deg$^{2}$)}}

\startdata
$B$   & Subaru 8.2~m & 0.96 & 27.8 & 0.27 & 0.20 \\
$B$   & CFHT 3.6~m   & 0.97 & 27.6 & 0.49 & 0.49 \\
$V$   & Subaru 8.2~m & 1.15 & 27.5 & 0.36 & 0.20 \\
$R$   & Subaru 8.2~m & 0.96 & 27.9 & 0.27 & 0.20 \\
$R$   & Subaru 8.2~m & 0.61 & 27.7 & 0.81 & 0.81 \\
$R$   & CFHT 3.6~m   & 0.89 & 27.9 & 0.49 & 0.49 \\
$I$   & Subaru 8.2~m & 1.30 & 26.4 & 0.36 & 0.20 \\
$z'$  & Subaru 8.2~m & 1.01 & 26.2 & 0.36 & 0.20 \\
$Z$   & CFHT 3.6m   & 0.95 & 26.3 & 0.49 & 0.49 
\enddata
\end{deluxetable}

\vfil\eject\clearpage

%
%
\begin{deluxetable}{lll}
\tablewidth{0pt}
  \tablecolumns{3}
\tablecaption{\label{spec_obs_summary}Summary of Spectroscopic
  Observations: Telescopes, Instruments, and Dates of Observations}
\tablehead{
  \colhead{Telescope} &
  \colhead{Instrument} &
\colhead{UT Dates}}
\startdata
 WIYN 3.5m & HYDRA & 2001 February 19-20 \\
 WIYN 3.5m & HYDRA & 2002 February 6-7 \\
 WIYN 3.5m & HYDRA & 2002 March 9-10 \\
 Keck II & DEIMOS & 2003 January 28-29 \\
 Keck II & DEIMOS & 2003 March 26 \\
 Keck II & DEIMOS & 2003 April 24-26 \\
 Keck II & DEIMOS & 2003 May 26 \\
 Keck II & DEIMOS & 2004 January 17-19 \\
 Keck II & DEIMOS & 2004 April 13-15 
\enddata
\end{deluxetable}

%
%
\begin{deluxetable}{lrcc}
\tablewidth{0pt}
  \tablecolumns{4}
  \tablecaption{\label{optical_soft}Median \emph{R}-band Magnitudes of the 0.5-2~keV CLASXS
    (combined) Sample} 
\tablehead{\colhead{$0.5-2$~keV Flux Range} &
           \colhead{} &
           \colhead{} &
           \colhead{} \\
           \colhead{($10^{-16}$ \flux) } &
           \colhead{Number} &
           \colhead{Median $0.5-2$~keV Flux} &
           \colhead{\emph{R}}}
\startdata
  1-3        & 36 (248)   & $2.2\times10^{-16}$ ($1.9\times10^{-16}$) & 24.7 (23.3) \\
  3-10      & 149 (325)   & $6.6\times10^{-16}$ ($6.1\times10^{-16}$) & 24.6 (23.9) \\
  10-30     & 190 (349)   & $1.9\times10^{-15}$ ($1.9\times10^{-15}$) & 23.4 (23.2) \\
  30-100     & 92 (188)   & $5.6\times10^{-15}$ ($5.6\times10^{-15}$) & 22.4 (22.2) \\
  100-300     & 37 (80)   & $1.8\times10^{-14}$ ($1.7\times10^{-14}$) & 20.6 (20.7) \\
  300-1000     & 6 (31)   & $4.6\times10^{-14}$ ($5.5\times10^{-14}$) & 20.6 (18.6)
\enddata
\end{deluxetable}

\vfil\eject\clearpage

%
%
\begin{deluxetable}{lrcc}
\tablewidth{0pt}
  \tablecolumns{4}
  \tablecaption{\label{optical_hard}Median \emph{R}-band Magnitudes of the $2-8$~keV CLASXS
    (Combined) Sample}
\tablehead{\colhead{$2-8$~keV Flux Range} &
           \colhead{Number} &
           \colhead{Median $2-8$~keV Flux} &
           \colhead{\emph{R}} \\
           \colhead{($10^{-16}$ \flux) } &
           \colhead{} &
           \colhead{} &
           \colhead{}}

\startdata
 1-3       & 6 (125)   & $2.5\times10^{-16}$ ($2.1\times10^{-16}$) & 24.4 (22.7) \\
 3-10      & 47 (258)  & $6.3\times10^{-16}$ ($6.1\times10^{-16}$) & 23.7 (23.7) \\
 10-30     & 98 (356)  & $2.0\times10^{-15}$ ($1.9\times10^{-15}$) & 24.1 (23.8) \\
 30-100    & 197 (369) & $5.9\times10^{-15}$ ($5.8\times10^{-15}$) & 24.0 (23.5) \\
 100-300   & 113 (236) & $1.7\times10^{-14}$ ($1.7\times10^{-14}$) & 22.5 (22.3) \\
 300-1000  & 18 (64)   & $5.4\times10^{-14}$ ($5.4\times10^{-14}$) & 21.2 (20.7) \\
 1000-3000 & 3 (70)    & $1.0\times10^{-13}$ ($1.8\times10^{-13}$) & 20.7 (18.0)
\enddata

\end{deluxetable}

%
%
\begin{deluxetable}{lrr}
\tablewidth{0pt}
  \tablecolumns{3}
  \tablecaption{\label{source_type}Number of \mbox{X-ray} Sources Per Spectral Type For
    Identified CLASXS Sources} 
\tablehead{\colhead{Class} & \colhead{Number} & \colhead{$\%$ of
    CLASXS Sample}}
\startdata
Stars & 20 & 4 \\
Star Formers & 73 & 14 \\
Broad-line AGNs & 106 & 20 \\
Seyferts & 44 & 8 \\
Absorbers & 28 & 5 
 \enddata

\end{deluxetable}

\end{document}